\documentclass[journal,twoside,web]{ieeecolor}
\usepackage{generic}
\usepackage{cite}
\usepackage{amsmath,amssymb,amsfonts}
\usepackage{algorithmic}
\usepackage{graphicx}
\usepackage{algorithm,algorithmic}
\usepackage{booktabs}
\usepackage{multirow}
\usepackage{hyperref}
\hypersetup{hidelinks}
\usepackage{textcomp}
\providecommand{\refname}{References}

\usepackage{etoolbox}
\makeatletter
\patchcmd{\thebibliography}{\itemsep 0pt plus pt\relax}
  {\itemsep 0pt plus 1pt\relax}{}{%
  \GenericWarning{}{ieeecolor itemsep patch did not apply}}
\patchcmd{\thebibliography}{\section*{\color{black}}}
  {\section*{\color{black}\refname}}{}{%
  \GenericWarning{}{ieeecolor refname patch did not apply}}
\makeatother
\def\BibTeX{{\rm B\kern-.05em{\sc i\kern-.025em b}\kern-.08em
    T\kern-.1667em\lower.7ex\hbox{E}\kern-.125emX}}

\newif\ifpreprint
\preprinttrue

\ifpreprint
  \makeatletter
  \def\ps@preprint{%
    \def\@oddfoot{}\def\@evenfoot{}%
    \def\@oddhead{\hbox{}\scriptsize\leftmark\hfil\thepage}%
    \def\@evenhead{\hbox{}\scriptsize\thepage\hfil\leftmark}}%
  \def\ps@titlepagestyle{%
    \def\@oddfoot{}\def\@evenfoot{}\def\@oddhead{}\def\@evenhead{}}%
  \makeatother
\else
\fi

\begin{document}

\bstctlcite{IEEEtran:BSTctl}

\title{Towards Whole-Study Screening for Congenital Heart Disease in Fetal Ultrasound Using Multiple Instance Learning}

\author{Mohamed~Azzam, Ruobing~Liu, Esther~C.~Ugwueke, Ziyang~Xu, Shibiao~Wan, Alex~Foy,\\
Abraham~Zabih, Jason~Christensen, Neil~Hamill, Ling~Li, and Jieqiong~Wang\\%
\thanks{\ifpreprint\else Manuscript submitted \today. \fi\textit{(Corresponding author: Jieqiong Wang.)}}%
\thanks{M. Azzam, R. Liu, E. C. Ugwueke, Z. Xu, and J. Wang are with the Department of Neurological Sciences, University of Nebraska Medical Center, Omaha, NE 68198 USA (e-mail: jiwang@unmc.edu).}%
\thanks{S. Wan is with the Department of Genetics, Cell Biology and Anatomy, University of Nebraska Medical Center, Omaha, NE 68198 USA.}%
\thanks{A. Foy, A. Zabih, J. Christensen, and L. Li are with the Department of Pediatrics, University of Nebraska Medical Center, Omaha, NE 68198 USA.}%
\thanks{N. Hamill is with the Department of Obstetrics and Gynecology, University of Nebraska Medical Center, Omaha, NE 68198 USA.}%
\thanks{Retrospective, de-identified imaging was analyzed under approved UNMC IRB protocol 0342-24-EP.}}

\maketitle

\begin{abstract}
Congenital heart disease (CHD) is the most common birth defect, yet a large fraction of cases remain undetected on prenatal ultrasound, in part because current artificial-intelligence methods assume that the key diagnostic frames have already been isolated from a study, by a clinician or by a view classifier. We remove that assumption and address CHD screening directly at the level of the whole ultrasound study. We propose a two-stage framework that first learns transferable frame representations by self-supervised masked-autoencoder pre-training on unlabeled fetal ultrasound, then identifies cardiac frames with a disease-robust module and aggregates them with a transformer-based multiple instance learning (MIL) model that produces a case-level diagnosis from study-level labels alone. The model further returns its highest-scoring frames for clinician review, and a hierarchical head separates critical from non-critical CHD. On the internal test set of our multi-source development cohort (FUSE), the proposed cardiac-gated MIL model reaches an area under the curve (AUC) of 0.985 with a specificity of 0.990, outperforming the reproduced NATMED ensemble (AUC 0.861, specificity 0.600) and the FetalCLIP foundation model (AUC 0.867, specificity 0.710). On an independent external cohort, all models initially perform near chance, but label-free CORAL adaptation raises the proposed model from an AUC of 0.513 to 0.944, whereas whole-study and view-dependent baselines do not recover. These results indicate that whole-study MIL with disease-robust cardiac-frame identification is an accurate and deployable route to prenatal CHD screening.
\end{abstract}

\begin{IEEEkeywords}
Congenital heart disease, fetal ultrasound, multiple instance learning, prenatal screening, domain adaptation, self-supervised learning.
\end{IEEEkeywords}

\section{Introduction}
\label{sec:introduction}

\IEEEPARstart{C}{ongenital} heart disease (CHD) affects approximately one in one hundred births and is the most common birth defect and a leading cause of infant mortality~\cite{Hoffman2002_incidence,GBD2017_chd}. In 2017, CHD caused more than 260{,}000 deaths worldwide, with the majority are infants~\cite{GBD2017_chd}. Critical forms, which require intervention in the first days of life, occur in roughly two to three per one thousand births. Therefore, effective prenatal screening and diagnosis are essential for identifying affected pregnancies, enabling referral to specialized fetal-cardiology care, facilitating delivery at appropriately equipped centers, and supporting timely postnatal intervention~\cite{Donofrio2014_aha}. To improve prenatal detection, international guidelines specify the cardiac views that should be obtained during a screening examination~\cite{Carvalho2013_isuog}. In practice, however, prenatal detection of CHD remains below fifty percent and varies widely across institutions~\cite{vanNisselrooij2020_missed,Quartermain2015_variation}. The detection rate is even worse in rural and underserved areas, where limited access to specialized expertise and adequate resources can further constrain prenatal CHD screening~\cite{Krishnan2021_disparities,Pinto2012_barriers,Chowdhury2024_disparities}. Closing this detection gap is the motivation for our work.

Ultrasound is widely used for fetal screening due to several practical advantages: it provides real-time imaging, is free of ionizing radiation, is relatively inexpensive, and is already integrated into routine obstetric care~\cite{Carvalho2013_isuog,Donofrio2014_aha}. Yet it remains challenging for automated analysis. A screening study typically contains hundreds to thousands of frames, but only a small fraction depict the fetal heart. Also, image quality varies with fetal position and operator expertise. Furthermore, the image appearance can shift across sites, scanners, and acquisition protocols. Therefore, any method intended for real-world CHD screening must analyze the entire study rather than rely on a small set of pre-selected cardiac images.

Artificial intelligence (AI), and deep learning in particular, has advanced rapidly across medical imaging, demonstrating strong performance in a wide range of image-analysis tasks~\cite{Campanella2019_wsimil}. Similar advances in fetal ultrasound show that clinically relevant anatomical and diagnostic information can be learned directly from ultrasound images~\cite{GarciaCanadilla2020_mlfetal,Baumgartner2017_sononet}. These developments motivate AI-assisted prenatal CHD screening. However, translating image-level performance to real-world screening requires models that can operate on complete studies, identify the informative cardiac frames, and produce reliable subject-level predictions.

\begin{figure}[!t]
\centering
\includegraphics[width=\columnwidth]{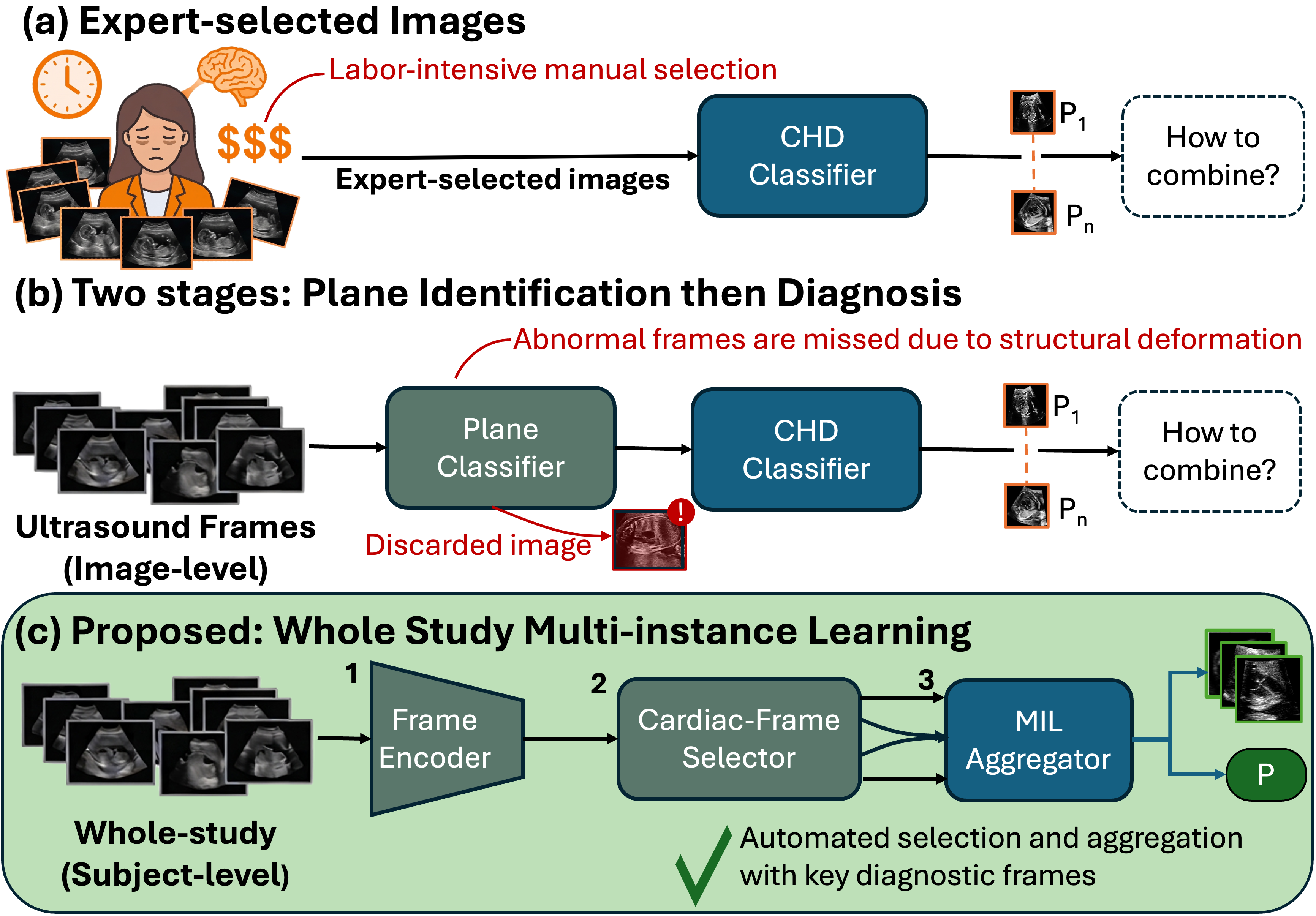}
\caption{Three designs for AI-based prenatal CHD screening. (a) Expert-selected images: a clinician manually selects representative frames, each is scored by a CHD classifier, and the per-image predictions must still be combined into a diagnosis. (b) Two-stage design: a plane classifier first identifies standard cardiac views before diagnosis. Because these classifiers are designed to recognize standard cardiac structures and are predominantly trained on normal anatomy, they may discard abnormal cardiac frames. Aggregation is again required as the design produces per-image predictions. (c) Proposed whole-study design: the entire study enters at subject level. The frame encoder embeds every frame, a disease-robust cardiac-frame identifier gates which frames enter the bag, and a MIL aggregator produces one subject-level prediction while surfacing the key diagnostic frames that support it.}
\label{fig:designs}
\end{figure}

Current AI pipelines do not fully satisfy these requirements. Existing approaches~\cite{Arnaout2021_natmed,Nurmaini2022_routine} commonly assume that diagnostically relevant cardiac frames have already been isolated, either manually or through a cardiac-plane identification step~\cite{Arnaout2021_natmed,Baumgartner2017_sononet}. Fig.~\ref{fig:designs}(a) and (b) illustrate these two prevailing designs. In the first, a clinician manually selects representative cardiac images that are subsequently analyzed by a CHD classifier. Although effective on curated images, this approach requires expert intervention and is labor intensive. The second design automates frame selection using a plane classifier to identify standard cardiac views before applying the CHD classifier. However, standard-plane classifiers are designed to recognize expected cardiac structures and are predominantly trained on normal anatomy. Structural abnormalities in CHD may therefore alter the anatomical patterns used for view recognition, causing diagnostically important abnormal cardiac frames to be discarded before they reach the CHD classifier~\cite{Tan2020_autochd}. Both designs also operate at the image level, producing separate predictions for individual frames that must subsequently be combined into a subject-level diagnosis. These limitations become particularly important in practical screening, where the relevant cardiac frames are not known in advance. In addition, severity classification for clinical triage remains limited, with only small studies reported~\cite{Nurmaini2022_routine}, and cross-site validation under differences in populations, disease prevalence, equipment, and acquisition protocols remains largely unexplored. A practical screening system should instead analyze the whole ultrasound study, identify informative cardiac frames without relying on normal standard-plane anatomy, aggregate information across those frames into a subject-level prediction, and remain reliable when transferred to a new clinical site.

Multiple instance learning (MIL) is a natural fit for this setting. MIL treats a study as a bag of instances (frames) with a single bag-level label, and learns both the bag prediction and the per-instance contributions without frame-level supervision~\cite{Dietterich1997_mil,Maron1997_mil}. Attention-based MIL makes the per-instance weights explicit~\cite{Ilse2018_attmil}, and MIL has become the standard tool for weakly supervised whole-slide pathology~\cite{Campanella2019_wsimil,Lu2021_clam}. These properties are what let a model trained only on study-level diagnoses nonetheless surface the individual frames that drive its decision, which is the behavior clinical review requires. Figure~\ref{fig:designs}(c) shows the resulting design: the whole study enters at the subject level, a learned cardiac gate replaces manual or view-based frame selection, and the MIL aggregator returns a single case-level prediction together with the ranked key diagnostic frames on which that prediction rests.

A screening tool must also generalize. Because ultrasound appearance shifts across scanners and clinics, a model trained at one site can degrade sharply at another. We therefore treat cross-site transfer as a first-class problem and study transfer to an independent external cohort, CARDIUM~\cite{Vega2025_cardium}, which was collected at a different institution and, unlike our development data, contributes curated cardiac frames rather than full sweeps, so it also tests whether a whole-study model accepts pre-selected input. To recover performance without target labels we use Deep CORAL~\cite{Sun2016_coral}, an unsupervised domain adaptation method based on correlation alignment (CORAL) that matches the second-order statistics of source and target features. This lets a deployed model be tuned to a new population from unlabeled studies alone, without additional annotation.

Prior AI work on fetal CHD has largely operated within the pre-selected-frame setting~\cite{GarciaCanadilla2020_mlfetal}. A prominent example is the ensemble of neural networks reported by Arnaout et al., hereafter referred to as NATMED, which detects complex CHD from standard cardiac views at expert level~\cite{Arnaout2021_natmed}. Other approaches include two-stage transfer-learning screens for duct-dependent lesions~\cite{Tang2023_twostage}, routine-screening detectors~\cite{Nurmaini2022_routine}, and disease-specific models for lesions such as hypoplastic left heart syndrome~\cite{Hlhs2023_ai}. FetalCLIP and related foundation models have further advanced representation learning for fetal ultrasound~\cite{Maani2026_fetalclip,Vislang2025_fetalus}. Despite these advances, existing methods largely rely on curated cardiac views rather than screening directly from the whole ultrasound study.

In this paper, we address whole-study CHD screening from fetal ultrasound. We propose a two-stage framework that pairs self-supervised representation learning with a disease-robust cardiac-frame identifier and a transformer-based MIL aggregator, train and test it on our multi-source development cohort (FUSE), and characterize its transfer to the independent external cohort CARDIUM~\cite{Vega2025_cardium} under label-free CORAL adaptation~\cite{Sun2016_coral}. The main contributions of this work are as follows.
\begin{enumerate}
\item We formulate CHD screening as a whole-study MIL problem and show that a transformer-based aggregator, applied to automatically identified cardiac frames, produces accurate case-level diagnoses from study-level labels alone while returning interpretable supporting frames.
\item We introduce a disease-robust cardiac-frame identification module that remains reliable on abnormal hearts, and we show that gating to identified cardiac frames is decisive for both accuracy and transfer.
\item We add a hierarchical severity head that separates critical from non-critical CHD, improving sensitivity to the under-represented non-critical class over a flat classifier.
\item We evaluate cross-site generalization on the independent external CARDIUM cohort~\cite{Vega2025_cardium} and show that label-free CORAL adaptation~\cite{Sun2016_coral} raises the transfer area under the receiver operating characteristic curve (AUC) of the proposed model from 0.513 to 0.944.
\end{enumerate}

The rest of this paper is organized as follows. Section~\ref{sec:related} reviews related work. Section~\ref{sec:methods} presents the proposed framework. Section~\ref{sec:experiments} reports the experimental setup and results. Section~\ref{sec:conclusion} concludes.

\section{Related Work}
\label{sec:related}

\subsection{Prenatal CHD diagnosis}
Automated fetal cardiac analysis has followed the clinical screening pipeline. View or standard-plane classifiers localize the cardiac sweep within a study~\cite{Baumgartner2017_sononet}, segmentation networks delineate chambers and vessels for biometry~\cite{Xu2020_dwnet}, and downstream models flag abnormality~\cite{Vsd2023_dl}. On top of this pipeline, several CHD-specific systems have been proposed: the NATMED ensemble of neural networks for complex CHD from standard views~\cite{Arnaout2021_natmed}, one-class adversarial screening of fetal echocardiograms~\cite{Gong2020_dgacnn}, two-stage transfer learning for duct-dependent lesions~\cite{Tang2023_twostage}, routine-screening detectors for major CHD~\cite{Nurmaini2022_routine}, and lesion-specific models for hypoplastic left heart syndrome~\cite{Hlhs2023_ai}. More recent work targets deployment-facing problems such as standard cardiac-cycle detection in video~\cite{Hfsccd2024}, clip extraction for remote expert review~\cite{Videoclip2025_chd}, and retrospective community imaging~\cite{Retro2024_chd}, and large fetal-ultrasound foundation models improve transferable representations~\cite{Maani2026_fetalclip,Vislang2025_fetalus,Fetalnet2025}. Comprehensive surveys document both the progress and the recurring gap~\cite{Review2024_aifetalecho,Review2025_aifetalped,Review2024_fetalusdl}: reported performance is high on curated views but is rarely established on whole studies or across sites. Our work targets that gap directly.

\subsection{Multiple-instance learning}
MIL learns a bag-level label from a set of instances without instance-level supervision~\cite{Dietterich1997_mil,Maron1997_mil}. Modern formulations replace fixed pooling with learned aggregation: attention-based pooling weights instances by relevance~\cite{Ilse2018_attmil}, and neural MIL architectures generalize these operators~\cite{Wang2018_revisitmil}. In computational pathology, MIL is the dominant paradigm for weakly supervised whole-slide classification, including clustering-constrained attention~\cite{Lu2021_clam}, dual-stream contrastive MIL~\cite{Li2021_dsmil}, transformer-based correlated MIL~\cite{Shao2021_transmil}, distribution-guided MIL~\cite{Qu2022_dgmil}, and intrinsically interpretable additive MIL~\cite{Javed2022_additivemil}. MIL has also been applied to volumetric and video medical data, including COVID-19 severity from CT~\cite{Han2021_covidmil}, ordinal grading~\cite{Ordinal2025_mil}, knowledge-driven prenatal abnormality classification~\cite{MedKnow2025_mil}, and multimodal echocardiographic diagnosis~\cite{Huang2025_ssmil}. Robustness of MIL under distribution shift has received specific attention~\cite{Zhang2020_stablemil,QGMIL2026}. These methods motivate our use of a transformer-based aggregator, but they operate on curated bags; we instead build the bag from a raw study by first identifying cardiac frames.

\subsection{Domain adaptation}
Distribution shift between training and deployment data is a central obstacle in medical imaging. Feature-alignment methods match source and target statistics, for example by aligning second-order statistics with CORAL~\cite{Sun2016_coral} or minimizing maximum mean discrepancy~\cite{Gretton2007_mmd}; adversarial methods learn domain-invariant features~\cite{Ganin2015_dann}; and cluster-alignment and self-training approaches refine the target decision boundary~\cite{Azzam2021_cluster}. In fetal ultrasound specifically, generalization across devices and clinics has been studied in low-resource settings~\cite{Sendra2023_africa} and multi-center benchmarks~\cite{Benchmark2026_fetalbiometry}, and dataset-level biases have been documented~\cite{Ethics2025_fetalus}. We adopt label-free CORAL as our primary adaptation mechanism because it requires no target annotations and aligns naturally with the whole-study feature representation.

\section{Methods}
\label{sec:methods}

\subsection{Problem formulation and overall framework}
\label{sec:framework}
We consider case-level CHD screening from a whole fetal ultrasound study. A study of subject $b$ yields an unordered set (bag) of $N_b$ B-mode frames, $\mathcal{B}_b=\{I_{b,1},\dots,I_{b,N_b}\}$, with a single study-level label $y_b\in\{0,1,2\}$, where $0$ denotes a healthy control (HC), $1$ a non-critical CHD (NCCHD), and $2$ a critical CHD (CCHD). Critical CHD denotes lesions that require catheter-based or surgical intervention in the neonatal period, typically duct-dependent circulations; non-critical CHD denotes structural disease that does not require neonatal intervention. Class assignment was made by the clinical collaborators from the confirmed postnatal diagnosis. The binary detection label is $\mathbf{1}[y_b>0]$, so detection and severity grading are nested tasks. No frame-level diagnostic annotation is available or required. Frame-level view annotations exist for a subset of subjects and are used only to train the cardiac-frame identifier and, where present, to construct training bags (Section~\ref{sec:cardiac}).

This is the multiple instance learning setting~\cite{Dietterich1997_mil,Maron1997_mil}: the subject is the bag, each frame is an instance, and supervision exists only at the bag level. Because the bag is a set, the bag-level scoring function must be permutation-invariant. The model factorizes as instance embedding, permutation-invariant aggregation, and bag classification,
\begin{equation}\label{eq:mil}
\begin{split}
F(\mathcal{B}_b) &= g\bigl(\Phi(\{\mathbf{h}_{b,i}\}_{i=1}^{N_b})\bigr),\qquad \mathbf{h}_{b,i}=f_\theta(I_{b,i}),\\
\Phi(\{\mathbf{h}_{b,\varsigma(i)}\}_{i=1}^{N_b}) &= \Phi(\{\mathbf{h}_{b,i}\}_{i=1}^{N_b}),
\end{split}
\end{equation}
for every permutation $\varsigma$ of $\{1,\dots,N_b\}$, where $f_\theta$ is a frozen self-supervised frame encoder producing embeddings $\mathbf{h}_{b,i}\in\mathbb{R}^{D}$ (Sections~\ref{sec:repr} and~\ref{sec:embed}), $\Phi$ is the learned aggregator (Section~\ref{sec:mil}), and $g$ is the bag classifier head.

\begin{figure*}[!t]
\centering
\includegraphics[width=0.85\textwidth]{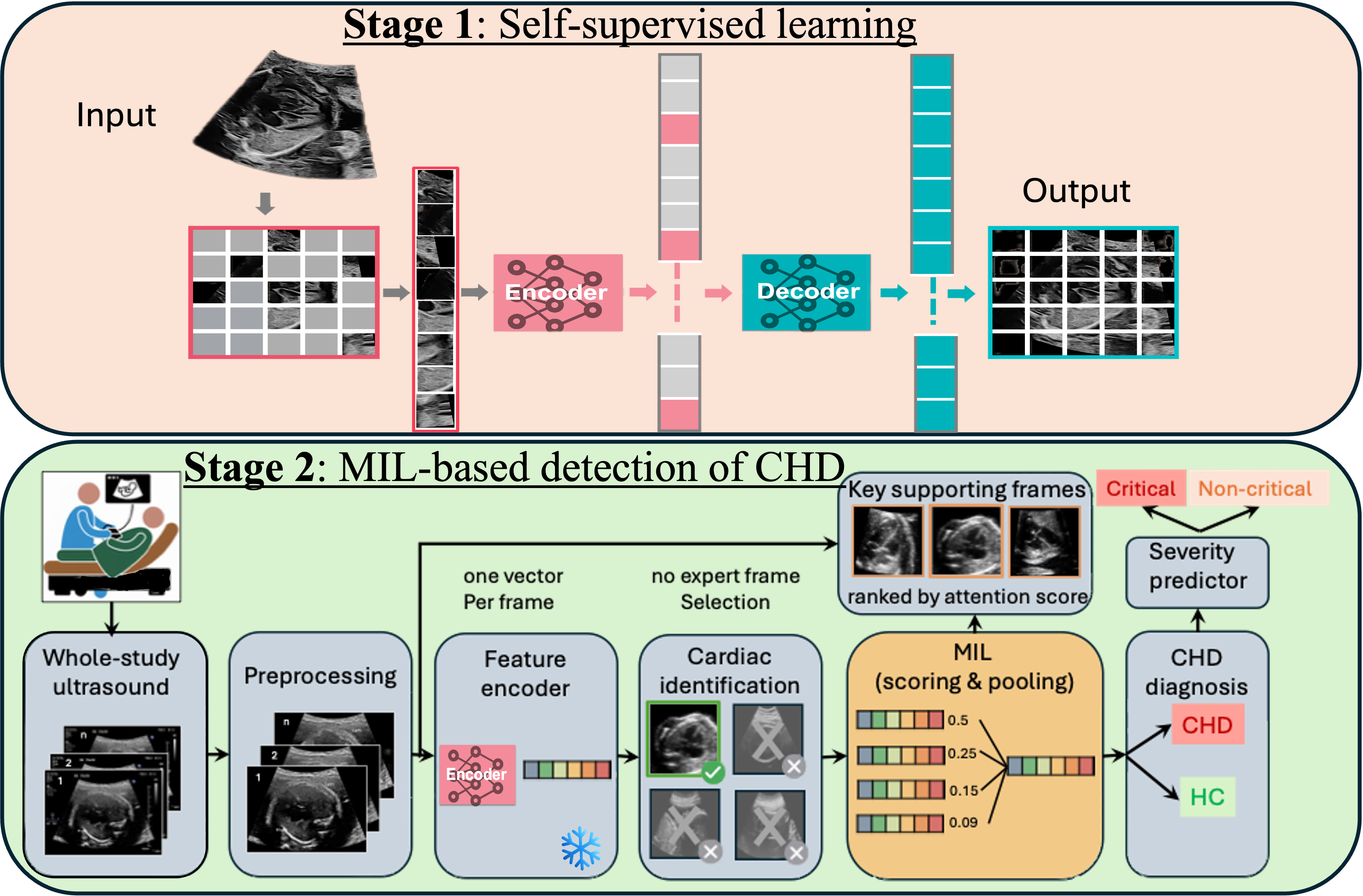}
\caption{Overview of the proposed two-stage framework. Stage~1: self-supervised pre-training of the frame encoder with a masked-autoencoder objective on unlabeled fetal ultrasound frames.
  Stage~2: MIL-based CHD detection. A whole-study ultrasound is preprocessed, every frame is embedded by the pre-trained encoder, which is kept frozen (snowflake) and yields one vector per frame; the cardiac-frame identification module passes only the features of the frames it identifies as cardiac, and a transformer aggregator with a classification token pools them into a study representation. A diagnosis head predicts CHD versus HC, a hierarchical severity head, trained as a separate branch, separates critical from non-critical CHD among predicted CHD cases. The attention from the classification token to each frame ranks the key supporting frames returned for clinician review. No expert frame selection is required at any point.}
\label{fig:pipeline}
\end{figure*}

Figure~\ref{fig:pipeline} shows the pipeline. Stage~1 pre-trains $f_\theta$ on unlabeled fetal ultrasound. In Stage~2 the encoder is frozen and embeds every retained frame of a study; a cardiac-frame identification module, trained on the same frozen embeddings, acts as a gate that passes forward only the frames it identifies as cardiac (Section~\ref{sec:cardiac}); the transformer-based MIL aggregator pools the gated bag through a learnable classification token whose attention over frames ranks the key supporting frames returned for clinician review (Section~\ref{sec:mil}); and task heads map the pooled representation to the outputs. The architecture supports two task heads over this shared backbone: a binary detection head
(Section~\ref{sec:adapt}), and a two-stage detect-then-grade head for severity (Section~\ref{sec:loss}).

Transfer from the FUSE development cohort to the external CARDIUM cohort (Section~\ref{sec:data}) involves three shifts, and each motivates a distinct design element. Appearance shift across scanners, operators, and protocols is addressed by second-order feature alignment (Section~\ref{sec:adapt}). Prior shift, from a CHD prevalence of about 41\% in the FUSE training pool to about 3\% in CARDIUM, is addressed by label-free operating-point rules at inference rather than in the loss (Section~\ref{sec:protocol}). Bag-composition shift is the largest of the three: a FUSE study is a full sweep of hundreds of frames (median 298 per training subject) spanning cardiac, abdominal, and non-diagnostic planes, whereas a CARDIUM study contributes a handful of curated, predominantly cardiac frames (median 4), so an aggregator trained to search a large heterogeneous bag would otherwise meet an input distribution it has never seen. Cardiac gating with a fixed frame cap aligns bag construction across cohorts (Section~\ref{sec:cardiac}).

\subsection{Representation learning (Stage 1)}
\label{sec:repr}
Diagnostic labels are scarce relative to the number of frames, so we learn frame features without them. The encoder $f_\theta$ is a Vision Transformer, ViT-L/16~\cite{Dosovitskiy2021_vit}, with $224\times224$ input, patch size 16 (196 patch tokens plus a class token), width 1024,  trained from random initialization on fetal ultrasound with the masked-autoencoder (MAE) objective~\cite{He2022_mae}. Pre-training used a leakage-controlled corpus of 369{,}846 masked B-mode frames from 1{,}650 fetuses, assembled from the two institutional cohorts and three public fetal-ultrasound datasets, where masked refers to the acquisition mask that blanks the region outside the ultrasound sector. Every subject reserved for diagnostic evaluation, namely the in-domain test set and the entire external cohort, was withheld from pre-training by construction, so no evaluation subject influenced the encoder. Public subjects without an annotated cardiac plane, which can never enter a diagnosis split, were retained as unlabeled anatomical variety.

Let $\mathbf{x}_j\in\mathbb{R}^{P}$ denote the $j$th of the 196 patches of the channel-normalized input, with $P=16\cdot16\cdot3=768$ values. A random subset $\mathcal{M}$ of patches is masked, with mask ratio $\rho=0.70$. The encoder then processes only the visible patches plus the class token, and a lightweight decoder reconstructs the masked patches from the encoder output, with a shared learnable mask token inserted at the removed positions. Reconstruction targets are per-patch standardized pixels and the loss is the mean squared error over masked patches only,
\begin{equation}\label{eq:mae}
\begin{split}
\mathcal{L}_{\mathrm{MAE}} &= \frac{1}{|\mathcal{M}|}\sum_{j\in\mathcal{M}}\frac{1}{P}\bigl\lVert \hat{\mathbf{x}}_j-\tilde{\mathbf{x}}_j\bigr\rVert_2^2,\\
\tilde{\mathbf{x}}_j &= \frac{\mathbf{x}_j-\mu_j\mathbf{1}}{\sqrt{\sigma_j^2+\varepsilon}},\\
\mu_j &= \tfrac{1}{P}\textstyle\sum_{p}x_{j,p},\qquad \sigma_j^2=\tfrac{1}{P-1}\textstyle\sum_{p}(x_{j,p}-\mu_j)^2,
\end{split}
\end{equation}
with $\hat{\mathbf{x}}_j$ the decoder prediction, $\varepsilon=10^{-6}$, and the unbiased variance of the reference implementation; visible patches contribute nothing to the loss. Augmentation reflects ultrasound acquisition variability rather than natural-image statistics (random resized crop, horizontal and vertical flips, a small random affine transform with zero fill matching the sector surround, and Gaussian blur), and channel normalization uses the corpus's own foreground-pixel statistics rather than ImageNet constants (Section~\ref{sec:config}). In-domain self-supervised features of this kind are expected to transfer better than supervised natural-image features because they are learned on fetal ultrasound and are not tied to the label distribution of any single task.

\subsection{Frozen frame embedding}
\label{sec:embed}
The pre-trained encoder is used strictly as a frozen feature extractor and is never fine-tuned for any task. Every downstream experiment is therefore a comparison of gating, aggregation, and adaptation strategies over one fixed representation, and the whole of Stage~2 trains in minutes on a single GPU. Each frame is resized to 256 pixels (bicubic), center-cropped to $224\times224$, and normalized with the same statistics as in pre-training, since a mismatch would shift every embedding. No masking is applied at inferencee, and the class token of the final encoder layer is the frame embedding,
\begin{equation}\label{eq:cls}
\begin{split}
\mathbf{U}^{(0)}_{b,i} &= \bigl[\mathbf{c}_{\mathrm{enc}}+\mathbf{E}^{\mathrm{pos}}_{0};\ \mathrm{PatchEmbed}(I_{b,i})+\mathbf{E}^{\mathrm{pos}}_{1:196}\bigr],\\
\mathbf{h}_{b,i} &= \mathrm{LN}\bigl(\mathrm{Block}_{24}\circ\cdots\circ\mathrm{Block}_{1}(\mathbf{U}^{(0)}_{b,i})\bigr)_{0}\in\mathbb{R}^{D},
\end{split}
\end{equation}
where $\mathbf{c}_{\mathrm{enc}}$ is the encoder class token, $\mathbf{E}^{\mathrm{pos}}$ the fixed positional embedding, and the subscript $0$ selects the class-token row. The MAE decoder is discarded. Extraction is a single deterministic pass per cohort over the frames retained by preprocessing (Section~\ref{sec:preproc}), with no augmentation.

\subsection{Cardiac-frame identification}
\label{sec:cardiac}
Multiple instance learning assumes that the instances of a bag bear on the bag label, each contributing evidence for or against it. A raw screening study violates that assumption. The sweep covers many organs, and a frame that shows no cardiac anatomy carries no evidence about a cardiac defect: it enters the aggregation as noise, it competes for the attention that belongs to the few diagnostic frames, and its appearance tracks the acquisition protocol rather than the diagnosis. The bag must therefore be cleaned before it is aggregated, and, because no frame-level annotation exists at deployment, it must be cleaned by the model itself.

We assume that view annotations are available for a subset of the training studies, drawn from both healthy and diseased cases, and use them to train a cardiac-frame identifier as a three-class frame classifier over non-target (NT), abdominal (ABDO), and cardiac frames, where cardiac is the union of the five target planes (apical four-chamber, three-vessel, three-vessel-trachea, left ventricular outflow tract, and right ventricular outflow tract). The classifier is a multilayer perceptron probe on the same frozen embedding used by the aggregator (layer normalization, then $1024\rightarrow256\rightarrow64\rightarrow3$ with GELU activations~\cite{Hendrycks2016_gelu} and dropout 0.3), calibrated by temperature scaling~\cite{Guo2017_calibration}, so gating costs one extra head rather than a second backbone. Training it on both normal and CHD studies, rather than on normal anatomy alone, improves cardiac-view recognition on CHD studies at no cost on normals; malformed hearts are exactly where a view classifier trained on normal anatomy fails, and those are the studies whose frames most need to be found.

The calibrated gate assigns every frame of every study a cardiac probability, and the bag is reduced to the frames whose probability clears a confidence threshold,
\begin{equation}\label{eq:gate}
\begin{split}
s_{b,i} &= \mathrm{softmax}\bigl(\phi(\mathbf{h}_{b,i})/T\bigr)_{\mathrm{card}},\\
\mathcal{B}^{\mathrm{card}}_b &= \bigl\{\, i\in\mathcal{A}_b \;:\; s_{b,i}\ \geq\ \tau_{\mathrm{card}} \,\bigr\},
\end{split}
\end{equation}
where $\phi$ is the probe, $T$ its fitted temperature, and $\mathcal{A}_b$ the frames of study $b$. The threshold $\tau_{\mathrm{card}}$ is a hyperparameter, fitted on a held-out annotated split and never on test or external data. 

\subsection{Transformer-based MIL aggregator (Stage 2)}
\label{sec:mil}
Attention-based MIL pooling~\cite{Ilse2018_attmil} scores each instance independently and takes a convex combination, so a frame's weight cannot depend on what else is in the bag. Diagnostic reading of a fetal study is not like that: a four-chamber view is informative relative to the outflow-tract views acquired in the same study, and an apparent abnormality is discounted when a better frame of the same plane contradicts it. We therefore aggregate with self-attention over the bag~\cite{Vaswani2017_attention,Shao2021_transmil}, which makes the contribution of each frame a function of the entire bag. The gated frame embeddings $\{\mathbf{h}_{b,i}\}_{i\in\mathcal{B}^{\mathrm{card}}_b}$, re-indexed $i=1,\dots,n_b$, enter the aggregator directly, without any learned projection, together with a learnable classification token $\mathbf{c}\in\mathbb{R}^{D}$; a shallow transformer encoder processes the sequence, and the transformed classification token is the bag representation,
\begin{subequations}\label{eq:agg}
\begin{align}
\mathbf{X}^{(0)}_b &= \bigl[\mathbf{c};\ \mathbf{h}_{b,1};\ \dots;\ \mathbf{h}_{b,N}\bigr]\in\mathbb{R}^{(N+1)\times D},\label{eq:agg-tokens}\\
\mathrm{head}_k(\mathbf{X}) &= \mathrm{softmax}\!\Bigl(\tfrac{(\mathbf{X}\mathbf{W}^{Q}_k)(\mathbf{X}\mathbf{W}^{K}_k)^{\!\top}}{\sqrt{d_k}}+\mathbf{M}_b\Bigr)\mathbf{X}\mathbf{W}^{V}_k,\label{eq:agg-mhsa}\\
\mathbf{z}_b &= \bigl[\mathrm{Enc}^{(L)}(\mathbf{X}^{(0)}_b;\mathbf{M}_b)\bigr]_{0}\in\mathbb{R}^{D},\qquad L=2,\label{eq:agg-cls}
\end{align}
\end{subequations}
where $n_b=|\mathcal{B}^{\mathrm{card}}_b|$ is the size of the gated bag, which varies from study to study, $N=\max_b n_b$ is the padded bag length within a batch, and $\mathbf{M}_b$ is an additive mask with $M_{b,ij}=0$ for valid key positions $j$ and $-\infty$ for padding; the class-token position is always valid, so no bag can be fully masked. $\mathrm{Enc}^{(L)}$ applies $L=2$ post-norm layers, each computing $\mathbf{X}\leftarrow\mathrm{LN}(\mathbf{X}+\mathrm{Drop}(\mathrm{MHSA}(\mathbf{X})))$ followed by $\mathbf{X}\leftarrow\mathrm{LN}(\mathbf{X}+\mathrm{Drop}(\mathrm{FFN}(\mathbf{X})))$, with four heads of dimension $d_k=256$ concatenated through an output projection, a narrow ReLU feed-forward bottleneck $1024\rightarrow128\rightarrow1024$, and dropout 0.5 on the attention and feed-forward paths. The feed-forward width is the main capacity constraint: with $d_{\mathrm{model}}=1024$, a conventional four-fold feed-forward layer would add about 8.4~M parameters per layer against 692 training bags. No positional encoding is used, so the aggregator is exactly permutation-equivariant over frame tokens and the class-token read-out is exactly permutation-invariant, the correct inductive bias for an unordered study. It remains dependent on bag composition, including bag size, which is the expressiveness the method claims and the reason bag construction must be aligned across cohorts. The bag classifier head maps $\mathbf{z}_b$ to logits through a compact multilayer perceptron,
\begin{equation}\label{eq:head}
\begin{split}
\mathbf{a}_b &= \mathrm{Drop}\bigl(\mathrm{ReLU}(\mathrm{LN}(\mathbf{W}_1\mathbf{z}_b+\mathbf{b}_1))\bigr)\in\mathbb{R}^{d_h},\quad d_h=64,\\
\mathbf{e}_b &= \mathrm{Drop}\bigl(\mathrm{ReLU}(\mathbf{W}_2\mathbf{a}_b+\mathbf{b}_2)\bigr)\in\mathbb{R}^{d},\quad d=32,\\
\boldsymbol{\ell}_b &= \mathbf{W}_3\mathbf{e}_b+\mathbf{b}_3\in\mathbb{R}^{C},
\end{split}
\end{equation}
where $C$ is the number of classes and $\mathbf{e}_b$ is the penultimate bag embedding at which domain alignment is applied (Section~\ref{sec:adapt}).

The key supporting frames returned for clinician review are ranked by the attention that the classification token pays to each frame in the last encoder layer, averaged over heads and renormalized over valid frames,
\begin{equation}\label{eq:keyframes}
a_{b,j}=\frac{m_{b,j}\,\bar{A}^{(L)}_{b,0,j}}{\sum_{k=1}^{N} m_{b,k}\,\bar{A}^{(L)}_{b,0,k}},\qquad
\bar{\mathbf{A}}^{(L)}_b=\frac{1}{4}\sum_{h=1}^{4}\mathbf{A}^{(L,h)}_b,
\end{equation}
where $m_{b,j}\in\{0,1\}$ marks valid frames, $\mathbf{A}^{(L,h)}_b$ is the attention matrix of head $h$ in layer $L$, row $0$ is the class-token query, and column $j$ is frame $j$; the class token's self-attention mass is discarded by the renormalization. Frames are ranked by Eq.~(\ref{eq:keyframes}) alone: the per-frame weights returned by the aggregator's default forward path come from a projection that is not on the loss path and carry no interpretive meaning. All attention figures in this paper use Eq.~(\ref{eq:keyframes}).

\subsection{Training objectives}
\label{sec:loss}
Class imbalance enters at three levels and is handled by three mechanisms. In this work, we use focal loss~\cite{Lin2017_focal} with inverse-frequency class weights. At the severity level, the two CHD subclasses are swamped by the detection task, and we use a hard CHD-masked grading term. 

\textbf{Focal loss.} For logits $\boldsymbol{\ell}$, true class $y$, class weights $\boldsymbol{\alpha}$, and focusing parameter $\gamma$, our implementation computes
\begin{equation}\label{eq:focal}
\begin{split}
\mathcal{L}_{\mathrm{FL}}(\boldsymbol{\ell},y;\boldsymbol{\alpha}) &= \bigl(1-p_y^{\,\alpha_y}\bigr)^{\gamma}\bigl(-\alpha_y\log p_y\bigr),\\
p_y &= \mathrm{softmax}(\boldsymbol{\ell})_y.
\end{split}
\end{equation}
The class weight enters the cross-entropy before the modulating factor is formed. We report the form that was actually optimized. Batch reduction is an unweighted mean, no label smoothing is used, and no resampling is used in the main training loop. Class weights are inverse class frequencies over subjects in the training partition, normalized to sum to the number of classes,
\begin{equation}\label{eq:classw}
\alpha_y = C\,\frac{1/n_y}{\sum_{k=1}^{C}1/n_k},\qquad \sum_{k=1}^{C}\alpha_k=C.
\end{equation}

\textbf{Two-stage detect-then-grade objective.} A flat three-class softmax over HC, NCCHD, and CCHD must learn a comparatively easy decision (is there disease) and a hard one (how severe) from one gradient. Because HC is the largest class and detection carries most of the loss, the two CHD subclasses collapse toward each other. We break this coupling with two parallel binary heads that read the same pooled representation $\mathbf{z}_b$, each one hidden layer deep ($1024\rightarrow64$, layer normalization, ReLU, dropout 0.5, $64\rightarrow2$): a detection head with logits $\boldsymbol{\ell}^{(1)}_b$ (HC versus CHD) and a grading head with logits $\boldsymbol{\ell}^{(2)}_b$ (NCCHD versus CCHD). With $\mathcal{I}$ the minibatch and $\mathcal{C}=\{b\in\mathcal{I}:y_b>0\}$ its CHD-positive bags,
\begin{equation}\label{eq:twostage}
\begin{split}
\mathcal{L}_{\mathrm{det\text{-}grade}} ={}& \frac{w_d}{|\mathcal{I}|}\sum_{b\in\mathcal{I}}\mathcal{L}_{\mathrm{FL}}\bigl(\boldsymbol{\ell}^{(1)}_b,\ \mathbf{1}[y_b>0];\ \boldsymbol{\alpha}_d\bigr)\\
&+\frac{w_g}{|\mathcal{C}|}\sum_{b\in\mathcal{C}}\mathcal{L}_{\mathrm{FL}}\bigl(\boldsymbol{\ell}^{(2)}_b,\ y_b-1;\ \mathbf{1}\bigr).
\end{split}
\end{equation}
Three properties are load-bearing. The grading term is a hard mask, not a weighting: HC bags contribute exactly zero gradient to the grading head. It is normalized by $|\mathcal{C}|$, not by the batch size, so its magnitude does not shrink as the HC fraction grows. It is defined as zero when a batch contains no CHD bag. We use the weighting factors $w_d$ and $w_g$ to maintain balanced subclass recall. The CHD detection term uses inverse-frequency weights $\boldsymbol{\alpha}_d$ to account for class imbalance in the training studies.

The two heads are combined by the chain rule, which is what makes the formulation hierarchical rather than two independent models,
\begin{equation}\label{eq:chain}
\begin{split}
P(\mathrm{CHD}\mid\mathcal{B}_b) &= \mathrm{softmax}(\boldsymbol{\ell}^{(1)}_b)_1,\\
P(\mathrm{HC}\mid\mathcal{B}_b) &= 1-P(\mathrm{CHD}\mid\mathcal{B}_b),\\
P(\mathrm{CCHD}\mid\mathcal{B}_b) &= P(\mathrm{CHD}\mid\mathcal{B}_b)\,\mathrm{softmax}(\boldsymbol{\ell}^{(2)}_b)_1,\\
P(\mathrm{NCCHD}\mid\mathcal{B}_b) &= P(\mathrm{CHD}\mid\mathcal{B}_b)\,\mathrm{softmax}(\boldsymbol{\ell}^{(2)}_b)_0.
\end{split}
\end{equation}
These probabilities sum to one by construction, and the binary CHD score used for detection metrics is $P(\mathrm{NCCHD}\mid\mathcal{B}_b)+P(\mathrm{CCHD}\mid\mathcal{B}_b)$ exactly. Severity is decoded with two thresholds so that detection and grading can be operated independently,
\begin{equation}\label{eq:grade}
\hat{y}_b=\begin{cases}
0, & P(\mathrm{CHD}\mid\mathcal{B}_b)\le\tau,\\
2, & P(\mathrm{CHD}\mid\mathcal{B}_b)>\tau\ \text{and}\ g_b\ge\tau_g,\\
1, & \text{otherwise},
\end{cases}
\end{equation}
where $g_b=\mathrm{softmax}(\boldsymbol{\ell}^{(2)}_b)_1=P(\mathrm{CCHD}\mid\mathrm{CHD},\mathcal{B}_b)$, the CHD detection threshold ($\tau=0.5$) and the severity threshold ($\tau_g$) are fitted on the validation-split subjects with $y_b>0$ against the target $\mathbf{1}[y_b=2]$.

\subsection{Cross-site adaptation}
\label{sec:adapt}
A model trained on one cohort (Section~\ref{sec:data}) must transfer to sites with different scanners, populations, and protocols. Our route is unsupervised domain adaptation with deep CORAL~\cite{Sun2016_coral}, a differentiable penalty that matches the second-order statistics of source and target representations and is optimized jointly with the supervised objective,
\begin{equation}\label{eq:coral}
\begin{split}
\mathcal{L}_{\mathrm{CORAL}} &= \frac{1}{4d^{2}}\bigl\lVert\boldsymbol{\Sigma}_S-\boldsymbol{\Sigma}_T\bigr\rVert_F^{2},\\
\boldsymbol{\Sigma}_S &= \frac{1}{n_S-1}\bigl(\mathbf{E}_S-\mathbf{1}\bar{\mathbf{e}}_S^{\top}\bigr)^{\!\top}\bigl(\mathbf{E}_S-\mathbf{1}\bar{\mathbf{e}}_S^{\top}\bigr),
\end{split}
\end{equation}
where $\mathbf{E}_S\in\mathbb{R}^{n_S\times d}$ and $\mathbf{E}_T\in\mathbb{R}^{n_T\times d}$ stack the penultimate bag embeddings $\mathbf{e}_b$ of Eq.~(\ref{eq:head}) for a labeled source batch and an unlabeled target batch, $\bar{\mathbf{e}}_S$ is the source batch mean, the covariance matrix $\boldsymbol{\Sigma}_T$ is defined analogously to $\boldsymbol{\Sigma}_S$, and the penalty is zero when either batch holds fewer than two bags. Alignment is applied at the 32-dimensional penultimate embedding rather than at the frame features or the logits: the frame encoder is frozen, so aligning frame features would not act on any trainable parameter, and aligning the two-dimensional logits would directly fight the classification objective, whereas the penultimate bag embedding is the last representation that is both trainable and task-relevant. The total objective is
\begin{equation}\label{eq:total}
\begin{split}
\mathcal{L} ={}& \frac{1}{n_S}\sum_{b=1}^{n_S}\mathcal{L}_{\mathrm{FL}}\bigl(\boldsymbol{\ell}^{s}_b,y^{s}_b;\boldsymbol{\alpha}\bigr)\\
&+\lambda\,\mathcal{L}_{\mathrm{CORAL}}(\mathbf{E}_S,\mathbf{E}_T),
\end{split}
\end{equation}
with $\lambda$ fixed at 10 throughout training. Each source batch is paired with a batch of unlabeled CARDIUM bags drawn from a separate cyclic loader and constructed using the same gate and threshold. CARDIUM labels are never used in any loss, metric, or threshold during training. 

\section{Experiments}
\label{sec:experiments}

\begin{table}[!t]
\centering
\footnotesize
\caption{Composition of the Stage-1 self-supervised pre-training corpus.}
\label{tab:pretrain}
\begin{tabular}{@{}lrr@{}}
\toprule
Source & Pre-training subjects & Frames \\
\midrule
In-house                                   & 419 & 363{,}182 \\
FETAL\_PLANES\_DB~\cite{Burgos2020_planes} & 947 &   4{,}674 \\
MFUSPAC~\cite{Sendra2023_africa}           & 115 &       402 \\
FASSD~\cite{DaCorreggio2023_fassd}         & 169 &   1{,}588 \\
\midrule
\textit{Total} & \textbf{1{,}650} & \textbf{369{,}846} \\
\bottomrule
\end{tabular}
\end{table}

\begin{table*}[t]
\centering
\footnotesize
\caption{Composition of the cohorts used for diagnosis training and evaluation.}
\label{tab:cohort}
\begin{tabular}{@{}llrrcccl@{}}
\toprule
 & & & & \multicolumn{2}{c}{CHD} & & \\
\cmidrule(lr){5-6}
Cohort & Source & $n$ & HC & Non-critical & Critical & View-annotated & Purpose \\
\midrule
\multirow{5}{*}{FUSE}
 & In-house                              & 527 & 171 & 145 & 211 & 26.0\% & \multirow{5}{*}{Training ($n=692$) and internal testing ($n=177$)} \\
 & FETAL\_PLANES\_DB~\cite{Burgos2020_planes} & 280 & 280 &   0 &   0 & 100\%  & \\
 & MFUSPAC~\cite{Sendra2023_africa}         &  61 &  61 &   0 &   0 & 100\%  & \\
 & RFCHD~\cite{Cai2023_china}               &   1 &   0 &   1 &   0 & 100\%  & \\
 & \textit{Total}                        & \textbf{869} & 512 & 146 & 211 & 55.1\% & \\
\midrule
External & CARDIUM~\cite{Vega2025_cardium} & 790 & 767 & \multicolumn{2}{c}{23 (2.9\%)} & 0\% & External test \\
\bottomrule
\end{tabular}
\end{table*}

\begin{figure*}[!t]
\centering
\includegraphics[width=0.7\textwidth]{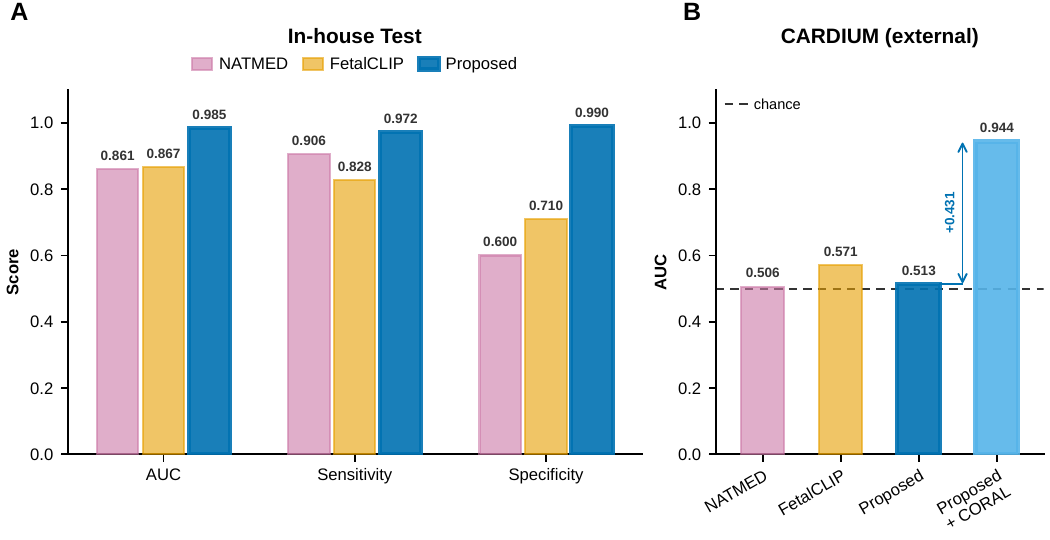}
\caption{CHD screening performance. (A) Internal FUSE test set (177 subjects): AUC, sensitivity, and specificity of NATMED, FetalCLIP, and the proposed cardiac-gated MIL model. (B) External CARDIUM cohort: AUC of NATMED, FetalCLIP, and the proposed model without adaptation, and of the proposed model after label-free CORAL adaptation. The dashed line marks chance.}
\label{fig:screening}
\end{figure*}

\begin{figure*}[!t]
\centering
\includegraphics[width=0.9\textwidth]{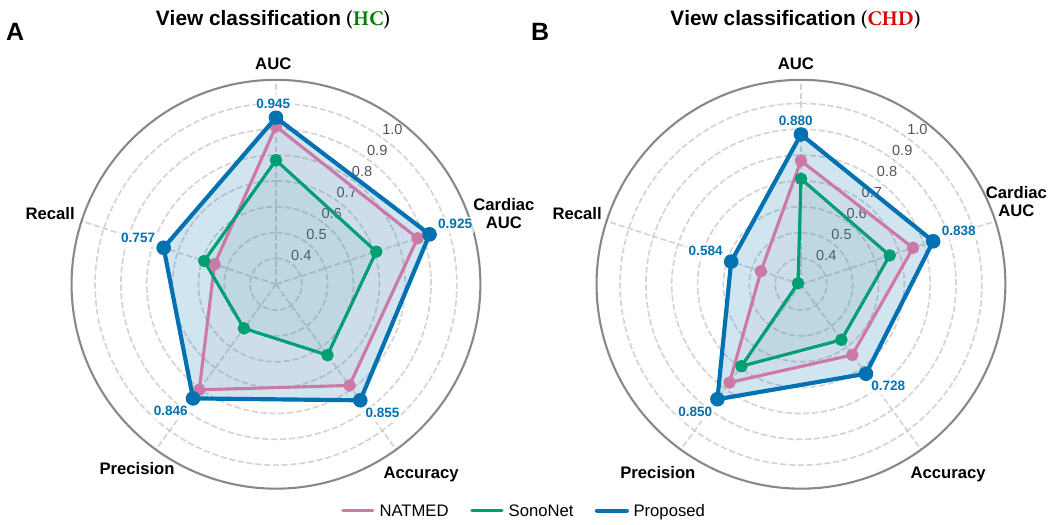}
\caption{Cardiac-view classification on the view-annotated internal test studies, for (A) HC and (B) CHD studies. Axes: AUC, cardiac-view AUC, precision, recall, and accuracy, for NATMED, SonoNet, and the proposed cardiac-frame identifier.}
\label{fig:viewcls}
\end{figure*}

\subsection{Datasets and cohort}
\label{sec:data}
Our development data form the FUSE cohort (fetal ultrasound multi-source development cohort), the union of an in-house cohort and three public fetal-ultrasound datasets, and we reserve an independent external cohort for out-of-domain evaluation, as reported in Table~\ref{tab:cohort}.

The in-house dataset was derived from the databases of Nebraska Medicine and Children's Nebraska hospitals, comprising ultrasound scans aggregated across seven medical institutions. The dataset spans three levels of care: routine prenatal screening, maternal-fetal medicine, and tertiary fetal cardiac care, corresponding to rural/community, secondary/referral, and tertiary sites, respectively. Ultrasound examinations were acquired using at least 13 scanners from Philips and GE, encompassing multiple scanner models across the participating institutions. Scans span gestational ages of 18 to 28 weeks, were collected between 2015 and 2024, and are 83\% urban and 17\% rural; the reference standard is postnatal echocardiographic evaluation.

The public sources are a European maternal-fetal plane dataset (FETAL\_PLANES\_DB)~\cite{Burgos2020_planes}, a five-country African low-resource dataset (MFUSPAC)~\cite{Sendra2023_africa}, and a Chinese recurrent-CHD family dataset (RFCHD)~\cite{Cai2023_china}. Public subjects enter the diagnosis splits only if they are evaluable for cardiac diagnosis, defined as having at least one frame annotated with a cardiac target plane at an acceptable quality. The remaining public subjects are used only for self-supervised pre-training.

FUSE comprises 869 subjects, which we split at the subject level into 692 training and 177 in-domain test subjects (approximately 80/20). Within the training subjects, a stratified 15\% subject-level validation split is  used for early stopping, checkpoint selection, and threshold fitting. A subset of 479 subjects, namely all 342 public subjects and 137 of the 527 in-house subjects (26.0\%), is annotated at the frame level by a team of pediatric cardiologists for view, one of the five cardiac target planes (apical four-chamber, three-vessel, three-vessel-trachea, and left and right ventricular outflow tract), abdominal, or non-target, and for image quality on a three-level scale (0 to 2). This annotated subset comprises 406 HC, 36 CCHD, and 37 NCCHD subjects; it covers 357 of the 692 training and 122 of the 177 test subjects.

The external cohort is CARDIUM~\cite{Vega2025_cardium}, whose CHD cases carry a reviewing cardiologist's diagnostic-confidence grade. After excluding color Doppler frames, the cohort holds 790 cases, including 767 HC and 23 CHD, corresponding to a CHD prevalence of 2.9\%. CARDIUM labels are never used for training, model selection, or threshold selection. Its unlabeled bags serve only as the alignment target of Section~\ref{sec:adapt}.

Table~\ref{tab:pretrain} decomposes the self-supervised pre-training corpus of Section~\ref{sec:config} by source. Every one of the 177 in-domain test subjects was withheld from it, as was the whole of CARDIUM, so no evaluation subject influenced the encoder. On the in-house side the pre-training subjects are exactly the 419 in-house training subjects; on the public side they also include the subjects that are not evaluable for cardiac diagnosis and therefore never enter a diagnosis split, which is why FETAL\_PLANES\_DB contributes 947 subjects to pre-training but 280 to the splits in Table~\ref{tab:cohort}. One further public source enters pre-training only: FASSD~\cite{DaCorreggio2023_fassd}, a fetal abdominal-structures dataset, holds abdominal-circumference images and no cardiac views, so none of its subjects is evaluable for cardiac diagnosis and it contributes abdominal anatomy to the encoder as unlabeled variety.

\subsection{Preprocessing}
\label{sec:preproc}

Preprocessing was applied consistently to the in-house, public, and CARDIUM cohorts. Each study was first de-identified, and scanner overlays, including text, measurements, and annotations, were removed. These overlays may contain scanner or site-specific information unrelated to cardiac anatomy and could encourage shortcut learning rather than reliance on clinically relevant image content~\cite{Lin2024_shortcut}. Each frame was then cropped to retain only the active ultrasound region, from which the $224\times224$ model input described in Section~\ref{sec:embed} was generated. Frames that were predominantly dark within the active ultrasound region were excluded.

For video clips, one frame was randomly sampled from every ten consecutive frames to reduce redundancy between adjacent frames while preserving coverage of the ultrasound sweep. The same preprocessing pipeline was used for all cohorts so that cross-site differences were not introduced by inconsistent data preparation.

\begin{table}[!t]
\centering
\caption{Internal CHD detection on 177 test studies (105 HC / 72 CHD). Best per column in bold.}
\label{tab:internal}
\begin{tabular}{@{}lccc@{}}
\toprule
Method & AUC & Sensitivity & Specificity \\
\midrule
NATMED~\cite{Arnaout2021_natmed} & 0.861 & 0.906 & 0.600 \\
FetalCLIP~\cite{Maani2026_fetalclip}            & 0.867 & 0.828 & 0.710 \\
\textbf{Proposed (cardiac-gated MIL)}       & \textbf{0.985} & \textbf{0.972} & \textbf{0.990} \\
\bottomrule
\end{tabular}
\end{table}

\subsection{Training configuration}
\label{sec:config}
\textbf{Stage 1: self-supervised pre-training.}
The frame encoder is a ViT-L/16~\cite{Dosovitskiy2021_vit} trained from random initialization with the MAE objective~\cite{He2022_mae} on $224\times224$ inputs. The pre-training corpus comprises 369{,}846 masked B-mode frames from 1{,}650 fetuses (see Table~\ref{tab:pretrain}). Input frames are masked at a ratio of 0.70 and reconstructed under a mean-squared error on per-patch normalized pixels. Augmentation comprises a random resized crop (scale 0.4 to 1.0), horizontal and vertical flips, a random affine transform (rotation $\pm10^\circ$, translation 10\%, shear $\pm5^\circ$), and Gaussian blur with probability 0.5, followed by normalization with foreground statistics computed on the corpus itself (mean 0.3529, standard deviation 0.2520) rather than ImageNet constants. We use AdamW~\cite{Loshchilov2019_adamw} with $\beta=(0.9,0.95)$ and weight decay 0.05, a base learning rate of $1.5\times10^{-4}$ scaled linearly with batch size to $7.03\times10^{-4}$ at an effective batch of 1200 frames on four NVIDIA L40S GPUs, 60 warm-up epochs followed by cosine decay over a 600-epoch schedule, bfloat16 precision, and gradient clipping at norm 1.0. The training took about 2.5 days. 

\textbf{Frozen extraction and cardiac-frame identifier.}
Each frame is resized to 256 pixels (bicubic), center-cropped to 224, and normalized with the pre-training statistics, and the final-layer class token computed without masking is the 1024-dimensional frame embedding. The cardiac-frame identifier, an MLP probe on this embedding (Section~\ref{sec:cardiac}), is trained with focal loss~\cite{Lin2017_focal} ($\gamma=2$, inverse-frequency class weights capped at 10), AdamW (learning rate $10^{-4}$, weight decay 0.05), batches of 1024 frames, and up to 100 epochs on 303 of the 357 view-annotated training subjects. The remaining 54 are held out for checkpoint selection, temperature scaling~\cite{Guo2017_calibration} ($T=0.516$), and threshold fitting.

\textbf{Stage 2: MIL aggregator, severity, and adaptation.}
The aggregator and heads are trained on the gated bags with AdamW (learning rate $10^{-4}$, weight decay 0.05), cosine annealing with warm restarts~\cite{Loshchilov2017_sgdr} ($T_0=20$, $T_{\mathrm{mult}}=2$, stepped once per epoch, so restarts fall at epochs 20 and 60), batches of 32 bags, gradient clipping at norm 1.0, up to 100 epochs, hidden width $d_h=64$, and dropout 0.5. Diagnosis uses focal loss with $\gamma=2$ and inverse-frequency class weights normalized to sum to the number of classes. A stratified 15\% subject-level validation split of the training pool, drawn with the run seed, drives early stopping on validation loss with patience 20, and the minimum-validation-loss checkpoint is retained. Severity training uses the two-stage objective of Section~\ref{sec:loss} with detection weight $w_d=1$, grading weight $w_g=2$, unit grade-class weights, and the grade threshold $\tau_g$ fitted on the validation split. CORAL adaptation adds the alignment penalty with a constant weight $\lambda=10$ and pairs each source batch with 32 unlabeled CARDIUM bags. Runs are seeded and reproducible up to data-loader worker nondeterminism. Because the encoder is frozen and each bag holds the extracted features instead of the raw image, one MIL run takes minutes on a single GPU.

\subsection{Evaluation protocol}
\label{sec:protocol}
For every run the minimum-validation-loss checkpoint is reloaded, and the in-domain test set and the CARDIUM cohort are then scored once. The area under the ROC curve (AUC) is threshold-free and is the primary metric. Sensitivity and specificity are also reported at the Youden-optimal threshold~\cite{Youden1950} fitted on the validation split. 

\subsection{Internal CHD detection}
\label{sec:internal}
On the internal test set of 177 studies (105 HC and 72 CHD), the proposed cardiac-gated MIL model outperforms both the reproduced NATMED ensemble~\cite{Arnaout2021_natmed} and the FetalCLIP foundation model~\cite{Maani2026_fetalclip}. Table~\ref{tab:internal} reports the comparison. The proposed model reaches an AUC of 0.985 and a specificity of 0.990, against 0.861 / 0.600 for NATMED and 0.867 / 0.710 for FetalCLIP. The contribution of cardiac gating itself is isolated in Section~\ref{sec:ablations}. Fig.~\ref{fig:screening}(A) plots the three metrics side by side.

\subsection{Cross-site domain adaptation}
\label{sec:cardium}
We evaluate transfer to the external CARDIUM cohort with and without label-free adaptation. Without adaptation, every trained model performs near chance, confirming that appearance shift, rather than a modeling failure, drives the drop. Label-free CORAL adaptation then recovers the proposed model dramatically, raising its AUC from 0.513 to 0.944, a gain of 0.431. In contrast, the view-dependent baselines improve only marginally: NATMED from 0.506 to 0.557 and FetalCLIP from 0.571 to 0.616. Table~\ref{tab:cardium} reports the comparison, and Fig.~\ref{fig:screening}(B) shows the unadapted AUCs of all three models together with the adapted proposed model. This is the central deployment result: restricting adaptation to identified cardiac frames is what makes label-free transfer succeed. Without the gate, non-cardiac content leaves the same model at an external AUC of 0.18 before adaptation (Table~\ref{tab:ablation}).

\begin{table}[!t]
\centering
\caption{External transfer to CARDIUM by model configuration, before and after label-free CORAL adaptation. Values are AUC.}
\label{tab:cardium}
\begin{tabular}{@{}llcc@{}}
\toprule
Model & Input & No adapt. & After adapt. \\
\midrule
NATMED~\cite{Arnaout2021_natmed} & Standard views & 0.506 & 0.557 \\
FetalCLIP~\cite{Maani2026_fetalclip} & Standard views & 0.571 & 0.616 \\
\textbf{Cardiac-gated MIL} & Cardiac frames & 0.513 & \textbf{0.944} \\
\bottomrule
\end{tabular}
\end{table}

\subsection{Representation quality}
\label{sec:reprquality}
\begin{figure}[!t]
\centering
\includegraphics[width=\columnwidth]{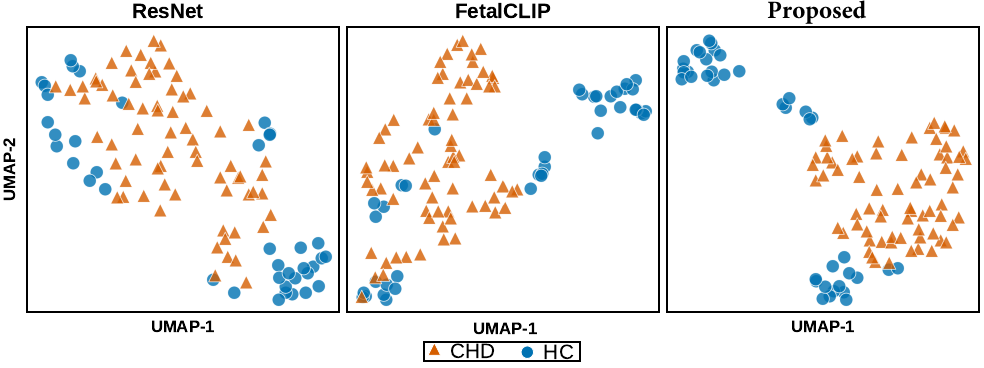}
\caption{UMAP projection of study-level representations of the in-house cohort, for a supervised ResNet, FetalCLIP, and the proposed encoder. Each point is one study, embedded as the average of the frame features of its apical four-chamber frames, and colored by diagnosis (CHD or HC).}
\label{fig:umap}
\end{figure}

Fig.~\ref{fig:umap} shows two-dimensional projections of the apical four-chamber representations from the in-house cohort for ResNet, FetalCLIP, and the proposed encoder using UMAP~\cite{McInnes2018_umap}. With the pretrained ResNet, HC and CHD studies are largely intermingled. FetalCLIP provides better organization, with partial separation of the healthy group, although substantial overlap remains. In contrast, the proposed encoder forms more distinct HC and CHD regions, with most of the remaining overlap concentrated near the class boundary. Importantly, this separation is obtained without diagnosis supervision: the encoder is frozen and the projection is computed directly from its learned features without using diagnosis labels. The resulting structure therefore reflects the quality of the learned fetal-ultrasound representation rather than the downstream MIL aggregator or diagnosis head, and is consistent with the performance advantage over FetalCLIP reported in Table~\ref{tab:internal}.

\subsection{Cardiac-view identification performance}
\label{sec:viewperf}

We evaluate the cardiac-frame identifier of Section~\ref{sec:cardiac} as a view classifier against the NATMED view classifier~\cite{Arnaout2021_natmed} and the standard-plane detector SonoNet~\cite{Baumgartner2017_sononet}. Our classifier outperforms both baselines on healthy and CHD studies (Fig.~\ref{fig:viewcls}), achieving a cardiac-view AUC of 0.925 on HC studies and 0.838 on CHD studies. Although recall decreases on CHD frames (0.584 versus 0.757 on HC studies), reflecting the greater difficulty of abnormal anatomy, the overall results indicate better generalization to abnormal cardiac frames than the compared methods.

\subsection{Hierarchical severity classification}
\label{sec:severity}
\begin{figure}[!t]
\centering
\includegraphics[width=\columnwidth]{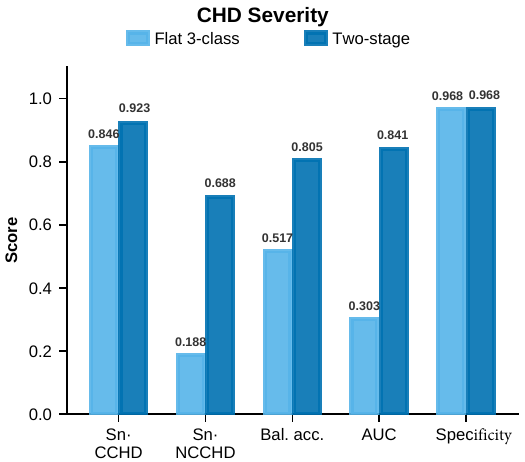}
\caption{Hierarchical severity classification on the internal test set, computed from calibrated probabilities. Sensitivity to CCHD and to NCCHD, balanced accuracy and AUC over the two CHD sub-classes, and specificity, for the flat three-class classifier and the two-stage hierarchical classifier.}
\label{fig:severity}
\end{figure}

Extending the model from HC-versus-CHD detection to three-class severity is harder: accuracy falls from 0.94 to 0.77 when the flat classifier must also separate critical from non-critical CHD. The two-stage hierarchical classifier recovers much of this loss where it matters most. It raises sensitivity to the non-critical class from 0.188 to 0.688 and sensitivity to the critical class from 0.846 to 0.923, while maintaining a specificity of 0.968. Because critical CHD carries the highest clinical cost if missed, the improvement in critical-class sensitivity at fixed specificity is the operationally important result. Over the two CHD sub-classes, balanced accuracy rises from 0.517 to 0.805 and AUC from 0.303 to 0.841 (Fig.~\ref{fig:severity}).

\subsection{Attention interpretability}
\label{sec:interp}
\begin{figure}[!t]
\centering
\includegraphics[width=\columnwidth]{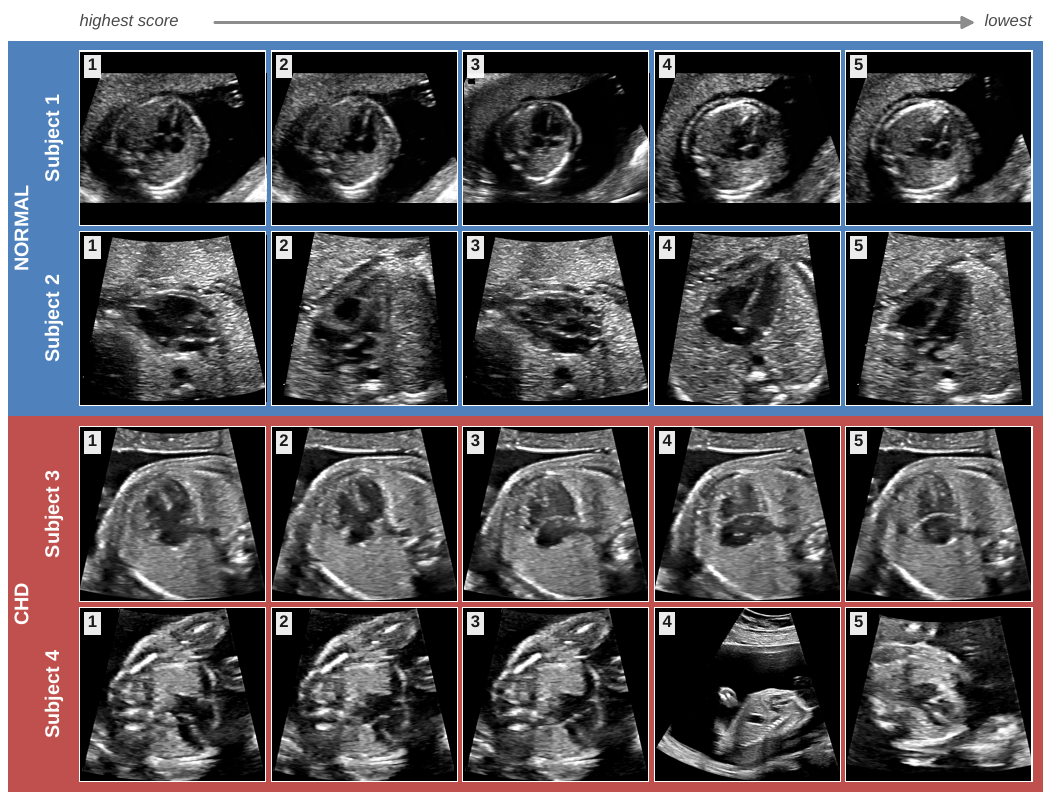}
\caption{Key supporting frames returned by the MIL aggregator. For two HC subjects (top, blue) and two CHD subjects (bottom, red), the five frames of the study that receive the most class-token attention in the last aggregator layer (Eq.~\ref{eq:keyframes}) are shown from left (highest attention) to right.}
\label{fig:interp}
\end{figure}

\begin{figure}[!t]
\centering
\includegraphics[width=\columnwidth]{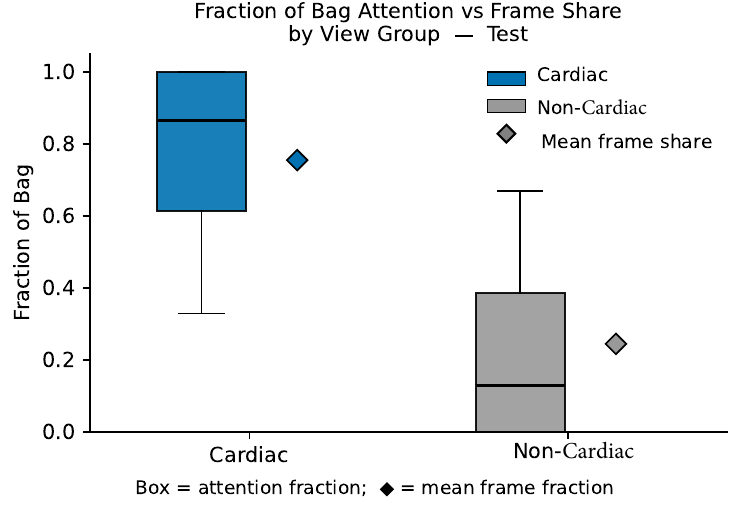}
\caption{Attention mass by view group on the 177 internal test subjects. Boxes show the fraction of each study's last-layer class-token attention (Eq.~\ref{eq:keyframes}) assigned to cardiac frames and to all non-cardiac frames (labeled non-cardiac in the plot, pooling the abdominal and non-target classes); diamonds mark the mean fraction of frames in each group.}
\label{fig:attmass}
\end{figure}

Because the aggregator ranks frames by last-layer class-token attention (Eq.~\ref{eq:keyframes}), we can compare the model's frame ranking against expert-identified key views. The top-ranked frames are standard cardiac views, predominantly the apical four-chamber view, at diagnostic quality, which indicates that the model locates the decisive plane without being told which frames matter. Fig.~\ref{fig:interp} shows the five highest-attention frames for two HC and two CHD subjects; in every case the top-ranked frames are cardiac views. The same preference holds in aggregate: across the view-annotated test studies, the fraction of a study's attention that falls on cardiac frames exceeds the fraction of frames that are cardiac, whereas non-cardiac frames receive less attention than their share of the bag (Fig.~\ref{fig:attmass}).

\subsection{Ablation study}
\label{sec:ablations}
Table~\ref{tab:ablation} isolates the contributions of whole-study MIL, cardiac-frame gating, and label-free CORAL adaptation on the internal test set and CARDIUM. In-domain, whole-study MIL alone achieves an AUC of 0.958, while cardiac-frame gating increases performance to 0.985. CORAL adaptation leaves the internal AUC unchanged at 0.985, indicating no loss in in-domain performance. The effect is more pronounced under cross-site transfer. Cardiac-frame gating improves the CARDIUM AUC from 0.180 to 0.513, and CORAL adaptation further increases it to 0.944, corresponding to gains of 0.333 and 0.431, respectively. Neither component is sufficient alone: gating improves frame selection but remains near chance without adaptation, whereas CORAL addresses the remaining domain shift in the learned representation. Together, the two components act at complementary stages of the pipeline: the gate determines which frames are analyzed, while CORAL aligns their feature distributions across domains.

\begin{table}[!t]
\centering
\caption{Ablation over the three components of the framework: MIL aggregation over the whole study, cardiac-frame gating, and label-free CORAL adaptation.}
\label{tab:ablation}
\begin{tabular}{@{}ccccc@{}}
\toprule
\multicolumn{3}{c}{Configuration} & \multicolumn{2}{c}{AUC} \\
\cmidrule(r){1-3} \cmidrule(l){4-5}
MIL & Cardiac gating & CORAL & Internal & CARDIUM \\
\midrule
\checkmark &            &            & 0.958 & 0.18  \\
\checkmark & \checkmark &            & 0.985 & 0.513 \\
\checkmark & \checkmark & \checkmark & 0.985 & 0.944 \\
\bottomrule
\end{tabular}
\end{table}

\section{Conclusion}
\label{sec:conclusion}
We presented a whole-study approach to prenatal CHD screening that removes the common assumption of clinician-preselected cardiac frames. The method pairs self-supervised representation learning with a disease-robust cardiac-frame identifier and a transformer-based MIL aggregator, produces case-level diagnoses and severity from study-level labels, and returns interpretable supporting frames. On the FUSE internal test set it reaches an AUC of 0.985 at a specificity of 0.990, and on an independent external cohort label-free CORAL adaptation restores near-random transfer to an AUC of 0.944, provided adaptation is restricted to identified cardiac frames. To our knowledge, this work represents a first attempt to directly address CHD screening from whole ultrasound studies and demonstrates that standard-plane identification can fail on abnormal cardiac images. Cross-site experiments further show that models trained on one dataset require adaptation before deployment at a new site because of differences in scanners, populations, disease prevalence, and acquisition protocols. Future work should investigate potential bias in cardiac-plane identification for abnormal hearts and extend CHD prediction beyond detection and severity to explicit disease-subtype identification.

\section*{Acknowledgment}
Research reported in this publication was supported by the National Institute of General Medical Sciences of the National Institutes of Health under Award Number P20GM152326, the Child Health Research Institute AI pilot, the University of Nebraska Collaboration Initiative Grant from the Nebraska Research Initiative (NRI). This project is also supported by the U.S. National Science Foundation under Award Numbers 2500836 and 2614824, the National Cancer Institute of the National Institutes of Health under Award Number R03CA317707, the National Institute on Alcohol Abuse and Alcoholism of the National Institutes of Health under Award Number R21AA032098. This research was supported by the State of Nebraska through the Pediatric Cancer Research Group, part of the Child Health Research Institute.  The content is solely the responsibility of the authors and does not necessarily represent the official views of the funding organizations.

\bibliographystyle{IEEEtran}
\bibliography{../Bibliography_base}

\begin{thebibliography}{10}
\providecommand{\url}[1]{#1}
\csname url@samestyle\endcsname
\providecommand{\newblock}{\relax}
\providecommand{\bibinfo}[2]{#2}
\providecommand{\BIBentrySTDinterwordspacing}{\spaceskip=0pt\relax}
\providecommand{\BIBentryALTinterwordstretchfactor}{4}
\providecommand{\BIBentryALTinterwordspacing}{\spaceskip=\fontdimen2\font plus
\BIBentryALTinterwordstretchfactor\fontdimen3\font minus
  \fontdimen4\font\relax}
\providecommand{\BIBforeignlanguage}[2]{{%
\expandafter\ifx\csname l@#1\endcsname\relax
\typeout{** WARNING: IEEEtran.bst: No hyphenation pattern has been}%
\typeout{** loaded for the language `#1'. Using the pattern for}%
\typeout{** the default language instead.}%
\else
\language=\csname l@#1\endcsname
\fi
#2}}
\providecommand{\BIBdecl}{\relax}
\BIBdecl

\bibitem{Hoffman2002_incidence}
J.~I.~E. Hoffman and S.~Kaplan, ``The incidence of congenital heart disease,''
  \emph{Journal of the American College of Cardiology}, vol.~39, no.~12, pp.
  1890--1900, 2002.

\bibitem{GBD2017_chd}
{GBD 2017 Congenital Heart Disease Collaborators}, ``Global, regional, and
  national burden of congenital heart disease, 1990-2017: a systematic analysis
  for the global burden of disease study 2017,'' \emph{The Lancet Child \&
  Adolescent Health}, vol.~4, no.~3, pp. 185--200, 2020.

\bibitem{Donofrio2014_aha}
M.~T. Donofrio, A.~J. Moon-Grady, L.~K. Hornberger, J.~A. Copel, M.~S.
  Sklansky, A.~Abuhamad, B.~F. Cuneo, J.~C. Huhta, R.~A. Jonas, A.~Krishnan,
  S.~Lacey, W.~Lee, E.~C. Michelfelder, G.~R. Rempel, N.~H. Silverman
  \emph{et~al.}, ``Diagnosis and treatment of fetal cardiac disease: a
  scientific statement from the {American Heart Association},''
  \emph{Circulation}, vol. 129, no.~21, pp. 2183--2242, 2014.

\bibitem{Carvalho2013_isuog}
J.~S. Carvalho, L.~D. Allan, R.~Chaoui, J.~A. Copel, G.~R. DeVore, K.~Hecher,
  W.~Lee, H.~Munoz, D.~Paladini, B.~Tutschek, and S.~Yagel, ``{ISUOG} practice
  guidelines (updated): sonographic screening examination of the fetal heart,''
  \emph{Ultrasound in Obstetrics \& Gynecology}, vol.~41, no.~3, pp. 348--359,
  2013.

\bibitem{vanNisselrooij2020_missed}
A.~E.~L. van Nisselrooij, A.~K.~K. Teunissen, S.~A. Clur, L.~Rozendaal,
  E.~Pajkrt, I.~H. Linskens, L.~Rammeloo, J.~M.~M. van Lith, N.~A. Blom, and
  M.~C. Haak, ``Why are congenital heart defects being missed?''
  \emph{Ultrasound in Obstetrics \& Gynecology}, vol.~55, no.~6, pp. 747--757,
  2020.

\bibitem{Quartermain2015_variation}
M.~D. Quartermain, S.~K. Pasquali, K.~D. Hill, D.~J. Goldberg, J.~C. Huhta,
  J.~P. Jacobs, M.~L. Jacobs, S.~Kim, and R.~M. Ungerleider, ``Variation in
  prenatal diagnosis of congenital heart disease in infants,''
  \emph{Pediatrics}, vol. 136, no.~2, pp. e378--e385, 2015.

\bibitem{Krishnan2021_disparities}
A.~Krishnan, M.~B. Jacobs, S.~A. Morris, S.~Peyvandi, A.~H. Bhat, A.~Chelliah,
  J.~S. Chiu, B.~F. Cuneo, G.~Freire, L.~K. Hornberger, L.~Howley, N.~Husain,
  C.~Ikemba, A.~Kavanaugh-McHugh, S.~Kutty, C.~Lee, K.~N. Lopez, A.~McBrien,
  E.~C. Michelfelder, N.~M. Pinto, R.~Schwartz, K.~W.~D. Stern, C.~Taylor,
  V.~Thakur, W.~Tworetzky, C.~Wittlieb-Weber, K.~Woldu, M.~T. Donofrio, and
  {Fetal Heart Society}, ``Impact of socioeconomic status, race and ethnicity,
  and geography on prenatal detection of hypoplastic left heart syndrome and
  transposition of the great arteries,'' \emph{Circulation}, vol. 143, no.~21,
  pp. 2049--2060, 2021.

\bibitem{Pinto2012_barriers}
N.~M. Pinto, H.~T. Keenan, L.~L. Minich, M.~D. Puchalski, M.~Heywood, and L.~D.
  Botto, ``Barriers to prenatal detection of congenital heart disease: a
  population-based study,'' \emph{Ultrasound in Obstetrics \& Gynecology},
  vol.~40, no.~4, pp. 418--425, 2012.

\bibitem{Chowdhury2024_disparities}
D.~Chowdhury, P.~A. Elliott, S.~Y. Asaki, S.~Amdani, Q.-T. Nguyen, C.~Ronai,
  S.~Tierney, V.~Y. Levy, K.~Puri, C.~A. Altman, J.~N. Johnson, and J.~S.
  Glickstein, ``Addressing disparities in pediatric congenital heart disease: a
  call for equitable health care,'' \emph{Journal of the American Heart
  Association}, vol.~13, no.~13, p. e032415, 2024.

\bibitem{Campanella2019_wsimil}
G.~Campanella, M.~G. Hanna, L.~Geneslaw, A.~Miraflor, V.~Werneck Krauss~Silva,
  K.~J. Busam, E.~Brogi, V.~E. Reuter, D.~S. Klimstra, and T.~J. Fuchs,
  ``Clinical-grade computational pathology using weakly supervised deep
  learning on whole slide images,'' \emph{Nature Medicine}, vol.~25, no.~8, pp.
  1301--1309, 2019.

\bibitem{GarciaCanadilla2020_mlfetal}
P.~Garcia-Canadilla, S.~Sanchez-Martinez, F.~Crispi, and B.~Bijnens, ``Machine
  learning in fetal cardiology: what to expect,'' \emph{Fetal Diagnosis and
  Therapy}, vol.~47, no.~5, pp. 363--372, 2020.

\bibitem{Baumgartner2017_sononet}
C.~F. Baumgartner, K.~Kamnitsas, J.~Matthew, T.~P. Fletcher, S.~Smith, L.~M.
  Koch, B.~Kainz, and D.~Rueckert, ``{SonoNet}: real-time detection and
  localisation of fetal standard scan planes in freehand ultrasound,''
  \emph{IEEE Transactions on Medical Imaging}, vol.~36, no.~11, pp. 2204--2215,
  2017.

\bibitem{Arnaout2021_natmed}
R.~Arnaout, L.~Curran, Y.~Zhao, J.~C. Levine, E.~Chinn, and A.~J. Moon-Grady,
  ``An ensemble of neural networks provides expert-level prenatal detection of
  complex congenital heart disease,'' \emph{Nature Medicine}, vol.~27, no.~5,
  pp. 882--891, 2021.

\bibitem{Nurmaini2022_routine}
S.~Nurmaini, R.~U. Partan, N.~Bernolian, A.~I. Sapitri, B.~Tutuko, M.~N.
  Rachmatullah, A.~Darmawahyuni, F.~Firdaus, and J.~C. Mose, ``Deep learning
  for improving the effectiveness of routine prenatal screening for major
  congenital heart diseases,'' \emph{Journal of Clinical Medicine}, vol.~11,
  no.~21, p. 6454, 2022.

\bibitem{Tan2020_autochd}
J.~Tan, A.~Au, Q.~Meng, S.~Finesilver-Smith, J.~Simpson, D.~Rueckert,
  R.~Razavi, T.~Day, D.~Lloyd, and B.~Kainz, ``Automated detection of
  congenital heart disease in fetal ultrasound screening,'' in \emph{Medical
  Ultrasound, and Preterm, Perinatal and Paediatric Image Analysis
  (ASMUS/PIPPI, MICCAI Workshops)}, ser. Lecture Notes in Computer Science,
  vol. 12437.\hskip 1em plus 0.5em minus 0.4em\relax Springer, 2020, pp.
  243--252.

\bibitem{Dietterich1997_mil}
T.~G. Dietterich, R.~H. Lathrop, and T.~Lozano-P\'erez, ``Solving the multiple
  instance problem with axis-parallel rectangles,'' \emph{Artificial
  Intelligence}, vol.~89, no. 1--2, pp. 31--71, 1997.

\bibitem{Maron1997_mil}
O.~Maron and T.~Lozano-P\'erez, ``A framework for multiple-instance learning,''
  \emph{Advances in Neural Information Processing Systems}, vol.~10, 1997.

\bibitem{Ilse2018_attmil}
M.~Ilse, J.~M. Tomczak, and M.~Welling, ``Attention-based deep multiple
  instance learning,'' in \emph{Proc. 35th Int. Conf. Machine Learning (ICML)},
  ser. Proc. Machine Learning Research, vol.~80, 2018, pp. 2127--2136.

\bibitem{Lu2021_clam}
M.~Y. Lu, D.~F.~K. Williamson, T.~Y. Chen, R.~J. Chen, M.~Barbieri, and
  F.~Mahmood, ``Data-efficient and weakly supervised computational pathology on
  whole-slide images,'' \emph{Nature Biomedical Engineering}, vol.~5, no.~6,
  pp. 555--570, 2021.

\bibitem{Vega2025_cardium}
D.~Vega, H.~V. Ceballos, J.~S. Vera, S.~Rodriguez, A.~Perez, A.~Castillo,
  M.~Escobar, D.~Londo\~no, L.~A. Sarmiento, C.~I. Castro, N.~Rodriguez, J.~C.
  Brice\~no, and P.~Arbelaez, ``{CARDIUM}: congenital anomaly recognition with
  diagnostic images and unified medical records,'' in \emph{Proc. IEEE/CVF Int.
  Conf. Computer Vision Workshops (ICCVW)}, 2025, pp. 1204--1213.

\bibitem{Sun2016_coral}
B.~Sun and K.~Saenko, ``Deep {CORAL}: correlation alignment for deep domain
  adaptation,'' in \emph{Computer Vision -- ECCV 2016 Workshops}, ser. Lecture
  Notes in Computer Science, vol. 9915, 2016, pp. 443--450.

\bibitem{Tang2023_twostage}
J.~Tang, Y.~Liang, Y.~Jiang, J.~Liu, R.~Zhang, D.~Huang, C.~Pang, C.~Huang,
  D.~Luo, X.~Zhou, R.~Li, K.~Zhang, B.~Xie, L.~Hu, F.~Zhu \emph{et~al.}, ``A
  multicenter study on two-stage transfer learning model for duct-dependent
  {CHDs} screening in fetal echocardiography,'' \emph{npj Digital Medicine},
  vol.~6, p. 143, 2023.

\bibitem{Hlhs2023_ai}
T.~G. Day, S.~Budd, J.~Tan, J.~Matthew, E.~Skelton, V.~Jowett, D.~Lloyd,
  A.~Gomez, J.~V. Hajnal, R.~Razavi, B.~Kainz, and J.~M. Simpson, ``Prenatal
  diagnosis of hypoplastic left heart syndrome on ultrasound using artificial
  intelligence: How does performance compare to a current screening
  programme?'' \emph{Prenatal Diagnosis}, vol.~44, no. 6-7, pp. 717--724, 2024.

\bibitem{Maani2026_fetalclip}
F.~Maani, N.~Saeed, T.~J. Saleem, Z.~Farooq, H.~Alasmawi, W.~Diehl,
  A.~Mohammad, G.~Waring, S.~Valappil, L.~Bricker, and M.~Yaqub, ``{FetalCLIP}:
  a visual-language foundation model for fetal ultrasound image analysis,''
  \emph{npj Digital Medicine}, 2026.

\bibitem{Vislang2025_fetalus}
X.~Guo, M.~Alsharid, H.~Zhao, Y.~Wang, J.~Lander, A.~T. Papageorghiou, and
  J.~A. Noble, ``A visually grounded language model for fetal ultrasound
  understanding,'' \emph{Nature Biomedical Engineering}, vol.~10, no.~8, pp.
  1629--1645, 2025.

\bibitem{Xu2020_dwnet}
L.~Xu, M.~Liu, Z.~Shen, H.~Wang, X.~Liu, X.~Wang, S.~Wang, T.~Li, S.~Yu,
  M.~Hou, J.~Guo, J.~Zhang, and Y.~He, ``{DW-Net}: A cascaded convolutional
  neural network for apical four-chamber view segmentation in fetal
  echocardiography,'' \emph{Computerized Medical Imaging and Graphics},
  vol.~80, p. 101690, 2020.

\bibitem{Vsd2023_dl}
Y.~Yang, B.~Wu, H.~Wu, W.~Xu, G.~Lyu, P.~Liu, and S.~He, ``Classification of
  normal and abnormal fetal heart ultrasound images and identification of
  ventricular septal defects based on deep learning,'' \emph{Journal of
  Perinatal Medicine}, vol.~51, no.~8, pp. 1052--1058, 2023.

\bibitem{Gong2020_dgacnn}
Y.~Gong, Y.~Zhang, H.~Zhu, J.~Lv, Q.~Cheng, H.~Zhang, Y.~He, and S.~Wang,
  ``Fetal congenital heart disease echocardiogram screening based on {DGACNN}:
  Adversarial one-class classification combined with video transfer learning,''
  \emph{IEEE Transactions on Medical Imaging}, vol.~39, no.~4, pp. 1206--1222,
  2020.

\bibitem{Hfsccd2024}
B.~Pu, K.~Li, J.~Chen, Y.~Lu, Q.~Zeng, J.~Yang, and S.~Li, ``{HFSCCD}: A hybrid
  neural network for fetal standard cardiac cycle detection in ultrasound
  videos,'' \emph{IEEE Journal of Biomedical and Health Informatics}, vol.~28,
  no.~5, pp. 2943--2954, 2024.

\bibitem{Videoclip2025_chd}
T.~G. Day, L.~Venturini, S.~F. Budd, A.~Farruggia, R.~Wright, J.~Matthew,
  V.~Zidere, T.~Vigneswaran, I.~Bo, A.~Savis, J.~Wolfenden, J.~Simpson,
  J.~Hajnal, B.~Kainz, and R.~Razavi, ``Video clip extraction from fetal
  ultrasound scans using artificial intelligence to allow remote second expert
  review for congenital heart disease,'' \emph{Prenatal Diagnosis}, vol.~45,
  no.~4, pp. 531--538, 2025.

\bibitem{Retro2024_chd}
C.~Athalye, A.~van Nisselrooij, S.~Rizvi, M.~C. Haak, A.~J. Moon-Grady, and
  R.~Arnaout, ``Deep-learning model for prenatal congenital heart disease
  screening generalizes to community setting and outperforms clinical
  detection,'' \emph{Ultrasound in Obstetrics \& Gynecology}, vol.~63, no.~1,
  pp. 44--52, 2024.

\bibitem{Fetalnet2025}
U.~Islam, Y.~A. Ali, M.~Al-Razgan, H.~Ullah, M.~A. Almaiah, Z.~Tariq, and K.~M.
  Wazir, ``{Fetal-Net}: enhancing maternal-fetal ultrasound interpretation
  through multi-scale convolutional neural networks and transformers,''
  \emph{Scientific Reports}, vol.~15, no.~1, p. 25665, 2025.

\bibitem{Review2024_aifetalecho}
J.~Zhang, S.~Xiao, Y.~Zhu, Z.~Zhang, H.~Cao, M.~Xie, and L.~Zhang, ``Advances
  in the application of artificial intelligence in fetal echocardiography,''
  \emph{Journal of the American Society of Echocardiography}, vol.~37, no.~5,
  pp. 550--561, 2024.

\bibitem{Review2025_aifetalped}
A.~Wang, T.~T. Doan, C.~Reddy, and P.-N. Jone, ``Artificial intelligence in
  fetal and pediatric echocardiography,'' \emph{Children}, vol.~12, no.~1,
  p.~14, 2025.

\bibitem{Review2024_fetalusdl}
M.~C. Fiorentino, F.~P. Villani, M.~Di~Cosmo, E.~Frontoni, and S.~Moccia, ``A
  review on deep-learning algorithms for fetal ultrasound-image analysis,''
  \emph{Medical Image Analysis}, vol.~83, p. 102629, 2023.

\bibitem{Wang2018_revisitmil}
X.~Wang, Y.~Yan, P.~Tang, X.~Bai, and W.~Liu, ``Revisiting multiple instance
  neural networks,'' \emph{Pattern Recognition}, vol.~74, pp. 15--24, 2018.

\bibitem{Li2021_dsmil}
B.~Li, Y.~Li, and K.~W. Eliceiri, ``Dual-stream multiple instance learning
  network for whole slide image classification with self-supervised contrastive
  learning,'' in \emph{Proc. IEEE/CVF Conf. Comput. Vis. Pattern Recognit.
  (CVPR)}, 2021, pp. 14\,313--14\,323.

\bibitem{Shao2021_transmil}
Z.~Shao, H.~Bian, Y.~Chen, Y.~Wang, J.~Zhang, X.~Ji, and Y.~Zhang,
  ``{TransMIL}: Transformer based correlated multiple instance learning for
  whole slide image classification,'' in \emph{Advances in Neural Information
  Processing Systems (NeurIPS)}, vol.~34, 2021, pp. 2136--2147.

\bibitem{Qu2022_dgmil}
L.~Qu, X.~Luo, S.~Liu, M.~Wang, and Z.~Song, ``{DGMIL}: Distribution guided
  multiple instance learning for whole slide image classification,'' in
  \emph{Medical Image Computing and Computer Assisted Intervention (MICCAI)},
  ser. Lecture Notes in Computer Science, vol. 13432.\hskip 1em plus 0.5em
  minus 0.4em\relax Springer, 2022, pp. 24--34.

\bibitem{Javed2022_additivemil}
S.~A. Javed, D.~Juyal, H.~Padigela, A.~Taylor-Weiner, L.~Yu, and A.~Prakash,
  ``Additive {MIL}: Intrinsically interpretable multiple instance learning for
  pathology,'' in \emph{Advances in Neural Information Processing Systems
  (NeurIPS)}, vol.~35, 2022.

\bibitem{Han2021_covidmil}
Z.~Li, W.~Zhao, F.~Shi, L.~Qi, X.~Xie, Y.~Wei, Z.~Ding, Y.~Gao, S.~Wu, J.~Liu,
  Y.~Shi, and D.~Shen, ``A novel multiple instance learning framework for
  {COVID-19} severity assessment via data augmentation and self-supervised
  learning,'' \emph{Medical Image Analysis}, vol.~69, p. 101978, 2021.

\bibitem{Ordinal2025_mil}
K.~Shiku, K.~Nishimura, D.~Suehiro, K.~Tanaka, and R.~Bise, ``Ordinal
  multiple-instance learning for ulcerative colitis severity estimation with
  selective aggregated transformer,'' in \emph{Proc. IEEE/CVF Winter Conf.
  Applications of Computer Vision (WACV)}, 2025.

\bibitem{MedKnow2025_mil}
H.~Liang, J.~Xu, Y.~Zhang, Y.~Huang, Y.~Zhang, X.~Yang, R.~Li, X.~Deng, Y.~Liu,
  G.~Tao, Y.~Wu, S.~Zhao, X.~Gao, and D.~Ni, ``Medical-knowledge driven
  multiple instance learning for classifying severe abdominal anomalies on
  prenatal ultrasound,'' in \emph{Medical Image Computing and Computer Assisted
  Intervention (MICCAI)}, ser. Lecture Notes in Computer Science.\hskip 1em
  plus 0.5em minus 0.4em\relax Springer, 2025, pp. 344--354.

\bibitem{Huang2025_ssmil}
Z.~Huang, X.~Yu, B.~S. Wessler, and M.~C. Hughes, ``Semi-supervised multimodal
  multi-instance learning for aortic stenosis diagnosis,'' in \emph{Proc. IEEE
  22nd Int. Symp. Biomedical Imaging (ISBI)}, 2025, pp. 1--5.

\bibitem{Zhang2020_stablemil}
W.~Zhang, L.~Liu, and J.~Li, ``Robust multi-instance learning with stable
  instances,'' in \emph{Proc. 24th European Conf. Artificial Intelligence
  (ECAI)}, ser. Frontiers in Artificial Intelligence and Applications, vol.
  325.\hskip 1em plus 0.5em minus 0.4em\relax IOS Press, 2020, pp. 1682--1689.

\bibitem{QGMIL2026}
L.~Zedda, D.~A. Mura, C.~Di~Ruberto, M.~Atzori, M.~F. Dasdelen, C.~Marr, and
  A.~Loddo, ``{QG-MIL}: A gated transformer aggregator for domain-agnostic
  multiple instance learning in medical imaging,'' in \emph{Medical Image
  Computing and Computer Assisted Intervention (MICCAI)}, ser. Lecture Notes in
  Computer Science.\hskip 1em plus 0.5em minus 0.4em\relax Springer, 2026, pp.
  440--450.

\bibitem{Gretton2007_mmd}
A.~Gretton, K.~M. Borgwardt, M.~Rasch, B.~Sch\"olkopf, and A.~J. Smola, ``A
  kernel method for the two-sample-problem,'' \emph{Advances in Neural
  Information Processing Systems}, vol.~19, pp. 513--520, 2007.

\bibitem{Ganin2015_dann}
Y.~Ganin and V.~Lempitsky, ``Unsupervised domain adaptation by
  backpropagation,'' in \emph{Proc. 32nd Int. Conf. Machine Learning (ICML)},
  ser. Proc. Machine Learning Research, vol.~37, 2015, pp. 1180--1189.

\bibitem{Azzam2021_cluster}
M.~Azzam, S.~Wu, A.~T. Gnanha, Q.~Jiao, and H.-S. Wong, ``Unsupervised domain
  adaptation via cluster alignment with maximum classifier discrepancy,'' in
  \emph{Proc. IEEE Int. Conf. Multimedia and Expo (ICME)}, 2021, pp. 1--6.

\bibitem{Sendra2023_africa}
C.~Sendra-Balcells, V.~M. Campello, J.~Torrents-Barrena, Y.~A. Ahmed,
  M.~Elattar, B.~Ohene-Botwe, P.~Nyangulu, W.~Stones, M.~Ammar, L.~N. Benamer,
  H.~N. Kisembo, S.~G. Sereke, S.~Z. Wanyonyi, M.~Temmerman, E.~Gratac\'os
  \emph{et~al.}, ``Generalisability of fetal ultrasound deep learning models to
  low-resource imaging settings in five {African} countries,'' \emph{Scientific
  Reports}, vol.~13, no.~1, p. 2728, 2023.

\bibitem{Benchmark2026_fetalbiometry}
C.~Di~Vece, Z.~Mao, N.~Avisdris, B.~Dromey, R.~Napolitano, D.~Ben~Bashat,
  F.~Vasconcelos, D.~Stoyanov, L.~Joskowicz, and S.~Bano, ``A multicentre
  benchmark dataset for comprehensive landmark-based fetal ultrasound
  biometry,'' \emph{Scientific Reports}, vol.~16, no.~1, p. 17405, 2026.

\bibitem{Ethics2025_fetalus}
M.~C. Fiorentino, S.~Moccia, M.~Di~Cosmo, E.~Frontoni, B.~Giovanola, and
  S.~Tiribelli, ``Uncovering ethical biases in publicly available fetal
  ultrasound datasets,'' \emph{npj Digital Medicine}, vol.~8, p. 355, 2025.

\bibitem{Dosovitskiy2021_vit}
A.~Dosovitskiy, L.~Beyer, A.~Kolesnikov, D.~Weissenborn, X.~Zhai,
  T.~Unterthiner, M.~Dehghani, M.~Minderer, G.~Heigold, S.~Gelly, J.~Uszkoreit,
  and N.~Houlsby, ``An image is worth 16x16 words: Transformers for image
  recognition at scale,'' in \emph{Int. Conf. Learning Representations (ICLR)},
  2021.

\bibitem{He2022_mae}
K.~He, X.~Chen, S.~Xie, Y.~Li, P.~Doll\'ar, and R.~Girshick, ``Masked
  autoencoders are scalable vision learners,'' in \emph{Proc. IEEE/CVF Conf.
  Computer Vision and Pattern Recognition (CVPR)}, 2022, pp. 15\,979--15\,988.

\bibitem{Hendrycks2016_gelu}
D.~Hendrycks and K.~Gimpel, ``Gaussian error linear units ({GELUs}),''
  arXiv:1606.08415, 2016.

\bibitem{Guo2017_calibration}
C.~Guo, G.~Pleiss, Y.~Sun, and K.~Q. Weinberger, ``On calibration of modern
  neural networks,'' in \emph{Proc. Int. Conf. Machine Learning (ICML)}, 2017,
  pp. 1321--1330.

\bibitem{Vaswani2017_attention}
A.~Vaswani, N.~Shazeer, N.~Parmar, J.~Uszkoreit, L.~Jones, A.~N. Gomez,
  L.~Kaiser, and I.~Polosukhin, ``Attention is all you need,'' \emph{Advances
  in Neural Information Processing Systems}, vol.~30, pp. 5998--6008, 2017.

\bibitem{Lin2017_focal}
T.-Y. Lin, P.~Goyal, R.~Girshick, K.~He, and P.~Doll{\'a}r, ``Focal loss for
  dense object detection,'' in \emph{Proc. IEEE Int. Conf. Computer Vision
  (ICCV)}, 2017, pp. 2980--2988.

\bibitem{Burgos2020_planes}
X.~P. Burgos-Artizzu, D.~Coronado-Guti\'errez, B.~Valenzuela-Alcaraz,
  E.~Bonet-Carne, E.~Eixarch, F.~Crispi, and E.~Gratac\'os, ``Evaluation of
  deep convolutional neural networks for automatic classification of common
  maternal fetal ultrasound planes,'' \emph{Scientific Reports}, vol.~10,
  no.~1, p. 10200, 2020.

\bibitem{DaCorreggio2023_fassd}
K.~S. Da~Correggio, R.~Noya~Galluzzo, L.~O. Santos, F.~Soares Muylaert~Barroso,
  T.~Zimmermann Loureiro~Chaves, A.~Sherlley Casimiro~Onofre, and A.~von
  Wangenheim, ``Fetal abdominal structures segmentation dataset using
  ultrasonic images,'' Mendeley Data, V1, 2023, doi: 10.17632/4gcpm9dsc3.1.

\bibitem{Cai2023_china}
R.~Cai, Y.~Tan, M.~Wang, H.~Yu, J.~Wang, Z.~Ren, Z.~Dong, Y.~He, Z.~Li, L.~Lin,
  and Y.~Gu, ``Detection of novel pathogenic variants in two families with
  recurrent fetal congenital heart defects,'' \emph{Pharmacogenomics and
  Personalized Medicine}, vol.~16, pp. 173--181, 2023.

\bibitem{Lin2024_shortcut}
M.~Lin, N.~Weng, K.~Mikolaj, Z.~Bashir, M.~B.~S. Svendsen, M.~G. Tolsgaard,
  A.~N. Christensen, and A.~Feragen, ``Shortcut learning in medical image
  segmentation,'' in \emph{Medical Image Computing and Computer Assisted
  Intervention (MICCAI)}, ser. Lecture Notes in Computer Science.\hskip 1em
  plus 0.5em minus 0.4em\relax Springer, 2024, pp. 623--633.

\bibitem{Loshchilov2019_adamw}
I.~Loshchilov and F.~Hutter, ``Decoupled weight decay regularization,'' in
  \emph{Int. Conf. Learning Representations (ICLR)}, 2019.

\bibitem{Loshchilov2017_sgdr}
I.~Loshchilov and F.~Hutter, ``{SGDR}: Stochastic gradient descent with warm
  restarts,'' in \emph{Int. Conf. Learning Representations (ICLR)}, 2017.

\bibitem{Youden1950}
W.~J. Youden, ``Index for rating diagnostic tests,'' \emph{Cancer}, vol.~3,
  no.~1, pp. 32--35, 1950.

\bibitem{McInnes2018_umap}
L.~McInnes, J.~Healy, and J.~Melville, ``{UMAP}: uniform manifold approximation
  and projection for dimension reduction,'' arXiv:1802.03426, 2018.

\end{thebibliography}

\end{document}